\documentclass[12pt]{article}
\usepackage{graphicx}
\usepackage{epsfig}
\usepackage{epsf,amsfonts,amssymb}
\newcommand{\be}{\begin{equation}}
\newcommand{\ee}{\end{equation}}
\newcommand{\bea}{\begin{eqnarray}}
\newcommand{\eea}{\end{eqnarray}}
\newcommand{\ba}{\begin{eqnarray}}
\newcommand{\ea}{\end{eqnarray}}
\newcommand{\nn}{\nonumber \\}
\newcommand{\eqn}[1]{(\ref{#1})}
\newcommand{\beq}{\begin{equation}}
\newcommand{\eeq}{\end{equation}}
\newcommand{\beqa}{\begin{eqnarray}}
\newcommand{\eeqa}{\end{eqnarray}}
\newcommand{\beqar}{\begin{eqnarray*}}
\newcommand{\eeqar}{\end{eqnarray*}}
\newcommand{\e}{\epsilon}
\newcommand{\reef}[1]{(\ref{#1})}

\newcommand{\ie}{{\it i.e.,}\ }

\newcommand{\ra}{\rightarrow}
\newcommand{\rom}[1]{{\mathrm{#1}}}
\newcommand{\sac}{\, , \qquad}

\newcommand{\bbe}[1]{\mbox{${\mathbb E}^{#1}$}}
\newcommand{\bbr}[1]{\mbox{${\mathbb R}^{#1}$}}
\newcommand{\bbc}[1]{\mbox{${\mathbb C}^{#1}$}}

\newcommand{\cala}{\mbox{${\cal A}$}}

\newcommand{\cale}{\mbox{${\cal E}$}}
\newcommand{\calf}{\mbox{${\cal F}$}}

\newcommand{\calh}{\mbox{${\cal H}$}}

\newcommand{\call}{\mbox{${\cal L}$}}

\newcommand{\calp}{\mbox{${\cal P}$}}

\begin{document}

\setlength{\unitlength}{1mm}

\thispagestyle{empty}
\rightline{\small hep-th/0408120}
\vspace*{2cm}

\begin{center}
{\bf \Large Supersymmetric black rings and three-charge supertubes}\\

\vspace*{1.4cm}

{\bf Henriette Elvang,}$^1\,$
{\bf Roberto Emparan,}$^2\,$
{\bf David Mateos,}$^3\,$
{\bf and Harvey S. Reall}$^4\,$

\vspace*{0.2cm}

{\it $^1\,$Department of Physics, University of California, Santa Barbara, CA 93106-9530, USA}\\[.3em]
{\it $^2\,$Instituci\'o Catalana de Recerca i Estudis Avan\c cats (ICREA),}\\
{\it Departament de F{\'\i}sica Fonamental, and}\\
{\it C.E.R. en Astrof\'{\i}sica, F\'{\i}sica de Part\'{\i}cules i Cosmologia,}\\
{\it Universitat de Barcelona, Diagonal 647, E-08028 Barcelona, Spain}\\[.3em]
{\it $^3\,$Perimeter Institute for Theoretical Physics, Waterloo,
Ontario N2J 2W9, Canada} \\
[.3em]
{\it $^4\,$Kavli Institute for Theoretical Physics, University of
  California, \\ Santa Barbara, CA 93106-4030, USA}

\vspace*{0.2cm}
{\tt elvang@physics.ucsb.edu, emparan@ub.edu,}\\
{\tt dmateos@perimeterinstitute.ca, reall@kitp.ucsb.edu}

\vspace*{0.5cm}
{NSF-KITP-04-95}

\vspace{.8cm} {\bf ABSTRACT}
\end{center}

We present supergravity solutions for 1/8-supersymmetric black
supertubes with three charges and three dipoles. Their reduction to
five dimensions yields supersymmetric black rings with
regular horizons and two independent angular momenta. The general
solution contains seven independent parameters and provides the
first example of non-uniqueness of supersymmetric black holes.
In ten dimensions, the solutions can be realized as D1-D5-P black
supertubes. We
also present a worldvolume construction of a supertube that
exhibits three dipoles explicitly. This description allows an arbitrary
cross-section but captures only one of the angular momenta.

\noindent

\vfill \setcounter{page}{0} \setcounter{footnote}{0}
\newpage

\tableofcontents

\setcounter{equation}{0}
\section{Introduction}

The black hole uniqueness theorems establish that, in four spacetime
dimensions, an equilibrium black hole has spherical topology and is
uniquely determined by its conserved charges. It was realized a few
years ago that these results do not extend to five dimensions. The
$D=5$ vacuum Einstein equations admit a solution describing a
stationary, asymptotically flat black hole with an event horizon of
topology $S^1 \times S^2$: a rotating {\it black ring} \cite{ER}.
The solution is {\it not} uniquely determined by its conserved charges
(mass and angular momentum) and, moreover, these charges do not even
distinguish black rings from black holes of spherical topology.
Charged black ring solutions with similar properties were
constructed in \cite{HE,EE}.

The black rings of \cite{ER,HE,EE} entail a finite violation of
black hole uniqueness, since there are finitely many solutions with
the same conserved charges. It has been suggested that black rings
exhibiting a continuously infinite violation of black hole uniqueness might also
exist \cite{reall:02}, and such solutions were recently constructed
\cite{RE}. In their simplest guise, these are described by solutions
of five-dimensional Einstein-Maxwell theory. Physically, they
describe rotating loops of magnetically charged black string. Since
the loop is contractible, these solutions carry no net magnetic
charge. They do carry, however, a non-zero magnetic dipole moment.
This is a non-conserved quantity, often referred to as `dipole
charge'. The black rings of \cite{RE} are characterized by their
mass, angular momentum and dipole charge, hence there is a
continuous infinity of solutions for fixed conserved charges.

The dipole charge has a simple microscopic interpretation
\cite{RE}. The black rings of \cite{RE} can be obtained from
dimensional reduction of an eleven-dimensional solution describing
M5-branes with four worldvolume directions wrapped on an internal
six-torus and one worldvolume direction forming the $S^1$ of the
black ring in the non-compact dimensions. The dipole charge of the
black ring is just the number of M5-branes present. The most general
solution of \cite{RE} has three independent dipole charges, since it
arises from the orthogonal intersection of three stacks of M5-branes
wrapped on $T^6$, with the common string of the intersection forming
the $S^1$ of the ring. Classically, the dipole charges are
continuous parameters, whereas in the quantum theory they are quantized
in terms of the number of branes in each stack.

Dipole moments and angular momentum play an important role in
another class of solutions of recent interest: the {\it supertubes}
of \cite{MT,EMT,MNT}. Black rings become black tubes when lifted to
higher dimensions, and ref.~\cite{EE} identified certain charged
non-supersymmetric black tubes as thermally excited states of
two-charge supertubes carrying D1- and D5-brane charges and a dipole
charge associated to a Kaluza-Klein monopole (KKM). Supertubes have
also been the subject of interest from a different direction
following the realization that the non-singular, horizon-free
supergravity solutions describing these objects are in one-to-one
correspondence with the Ramond-sector ground states of the
supersymmetric D1-D5 string intersection
\cite{mathur1,mathurdual,LMM}. It has
been conjectured that supergravity solutions for three-charge
supertubes might similarly account for the microstates of
supersymmetric five-dimensional black holes \cite{mathur2}. This
proposal has motivated a number of interesting studies on the D1-D5
system and supertubes \cite{sm,lunin,benakraus,bena,othertubes}
including the first examples of non-singular three-charge
supergravity supertubes without horizons \cite{sm,lunin}.

Investigations of the relationship between black rings and supertubes
have previously been done in the framework of the supergravity solutions
found in \cite{ER,HE,EE}. However, these do not admit a supersymmetric
limit with an event horizon, and this complicates understanding the
microscopic origin of their entropy.\footnote{Ref.~\cite{RE} made some
progress in this direction by studying a non-supersymmetric extremal
ring with a horizon.} It has been conjectured, though, that
supersymmetric black rings should exist \cite{benakraus,bena}. The
additional ingredient of supersymmetry of the black ring is important
for two reasons. First, many of the solutions of \cite{ER,HE,EE,RE} are
believed to be classically unstable, whereas a supersymmetric black ring
should be stable. Second, it should facilitate a precise quantitative
comparison between black rings, worldvolume supertubes, and the microscopic
conformal field theory of the D1-D5 system.

Recently, we found the first example of a supersymmetric black ring
\cite{EEMR}. It is a three-parameter solution of minimal $D=5$
supergravity. We shall see that, upon oxidation to ten dimensions,
this solution describes a black supertube carrying equal D1-brane,
D5-brane and momentum (P) charges, and equal D1, D5 and KKM dipole
moments. One purpose of the present paper is to generalize this
solution to allow for unequal charges and unequal dipole moments. We
shall present a seven-parameter black supertube solution labelled by
three charges, three dipole moments and the radius of the ring.

Our solution contains several previously known families of solutions
as special cases. First, it reduces to the solution of \cite{EEMR}
in the special case of three equal charges and three equal dipoles.
Second, it reduces to the two-charge supergravity supertubes of
\cite{EMT} when one of the charges and two of the dipoles
vanish. Third, in the zero-radius limit, the solution reduces to the
four-parameter solution describing supersymmetric black holes of
spherical topology \cite{bmpv}. Finally, in the infinite-radius
limit the dipole moments become conserved charges and the solution
reduces to the six-charge black string of \cite{bena}.

In five dimensions, our solution describes a supersymmetric black
ring. Although it is determined by seven parameters, it carries only
five independent conserved charges, namely the D1, D5 and momentum
charges (which determine the mass through the saturated BPS bound),
and two independent angular momenta. Hence classically, the
continuous violation of black hole uniqueness discovered for the
dipole black rings of \cite{RE} also extends to supersymmetric black
holes. In string/M-theory, the net charges and dipole charges must
be integer-quantized, since they represent the number of branes and
units of momenta. As a consequence of the charge quantization, the
violation of uniqueness is finite.

One might wonder whether this lack of uniqueness could be a problem for
a string theory calculation of the entropy of black rings. After all,
the original entropy calculations \cite{stva} simply counted all
microstates with the same conserved charges as the black hole, which
clearly will not work here. But note that there is no conflict with the
computation of the entropy of black holes of spherical topology, as
performed by Breckenridge, Myers, Peet and Vafa (BMPV) \cite{bmpv},
since the supersymmetric BMPV black hole has two equal angular momenta,
whereas our black rings always have unequal angular momenta. For the
rings themselves, the proposal of \cite{EE} is essentially that we
should resolve the non-uniqueness by counting only microstates belonging
to specific sectors of the D1-D5 CFT, with the precise sector being
determined by the values of the dipole charges. It will be interesting
to see whether this can be done at the orbifold point of the CFT. We
will make a few more comments on the issue of non-uniqueness in the
conclusions of the paper.

Two-charge supertubes were originally discovered as solutions of the
Dirac-Born-Infeld (DBI) effective action of a D-brane in a Minkowski
vacuum \cite{MT}. In this worldvolume picture the
branes associated to net charges are represented by fluxes on the
worldvolume of a tubular, higher-dimensional brane; the latter
carries no net charge itself but only a dipole charge. In this
description the back-reaction on spacetime of the branes is
neglected. The supergravity solution for a two-charge supertube
\cite{EMT,MNT} describes this backreaction.

The worldvolume description has proven extremely illuminating for
the physics of two-charge supertubes. For example, it has led to a
new way of counting the entropy of the D1-D5 system that does not
use its CFT description \cite{othertubes}. It is therefore desirable
to have an analogous description for three-charge supertubes. A
first step in this direction was given in
\cite{benakraus}, where a worldvolume description based on the
DBI action of a D6-brane that exhibits explicitly three charges and
two dipoles was found. However, generic three-charge supertubes
carry three dipoles, as can be understood from the fact each pair of
charges expands to a higher-dimensional brane. A worldvolume
description based on D-branes that incorporates the third dipole
seems problematic, since the latter necessarily corresponds to an
object that cannot be captured by an open string description, such
as NS5-branes or KKMs \cite{benakraus}. This difficulty can be
circumvented by going to M-theory, where the three branes with net
charges can be taken to be three orthogonal M2-branes, whereas the
three dipoles are associated to three M5-branes. (This is also the
most symmetric realization of the three-charge supertube.) We will
show that there exist supersymmetric  solutions of the effective
action of a single M5-brane in the M-theory Minkowski vacuum that
carry up to four M2-brane charges and six M5-brane dipoles. We call
these `calibrated supertubes' because the worldspace of the M5-brane
takes the form $S \times C$, where $S$ is a calibrated surface and
$C$ is an arbitrary curve. While the three-charge calibrated
supertube captures all dipoles and shows that an arbitrary
cross-section is possible, it also suffers from limitations. We will
discuss these in detail in the corresponding section. Suffice it to
say here that the calibrated supertube only captures one of the
angular momenta, as opposed to the two present in the supergravity
description.

The paper is organized as follows. In section \ref{sec:gensol} we
present the black supertube solution as an M-theory configuration
with three orthogonal M2-brane charges and three M5-branes
intersecting over a ring. Then in section \ref{sec:phys} we describe
other useful coordinates for the solution, calculate its physical
parameters, and study its causal structure and horizon geometry. In
section \ref{sec:d1d5p} we dualize the solution to a D1-D5-P black
supertube, which is shown to possess a remarkably rich structure.
Section \ref{sec:nonu} discusses how our black rings contain two
independent continuous parameters which are not fixed by the
asymptotic charges, and therefore realize infinite non-uniqueness of
supersymmetric black holes. In section \ref{sec:cases} we analyze
some particular cases contained within our general solution, and
study the `decoupling limit', relevant to AdS/CFT duality. Section
\ref{sec:worldv} is devoted to the construction of worldvolume
supertubes with three charges and three dipoles. In section
\ref{sec:sugravsworld} we give a preliminary comparison of the
supergravity black tubes with worldvolume supertubes. We conclude in
section \ref{sec:concl}.

The derivation and analysis of the solutions entail many technical
details that, for the sake of readability, we have found convenient
to move out of the main body of the paper into a number of extended
appendices. These are the derivation of the supersymmetric rings in
minimal supergravity and in $U(1)^N$ supergravity theories
(appendices \ref{app:minsugra} and \ref{app:U1N}), the conditions
for the absence of causal anomalies (appendix \ref{app:noctcs}), and
the proof of regularity of the horizon (appendix \ref{app:through}).

\bigskip

N.B.\ Following our publication of the minimal supersymmetric black 
ring in \cite{EEMR}, the solutions describing three charge
supersymmetric black rings have been found independently by two other
groups \cite{BW,GG2}. Solutions describing concentric black rings have
also been constructed \cite{JGJG,GG2}.

\setcounter{equation}{0}
\section{Three-charge black supertube in M-theory}
\label{sec:gensol}

The most symmetric realization of a supertube with three charges and
three dipoles is an M-theory configuration consisting of three
M2-branes and three M5-branes oriented as indicated by the array\footnote{
In such arrays, we shall reserve capital letters (M2) for branes
carrying conserved charges and lower case letters (m5) for branes
carrying dipole charges.}
\be
\begin{array}{rcccccccl}
Q_1 \,\, \mbox{M2:}\,\,\, & 1 & 2 & \_ & \_ & \_ & \_ & \_ & \, \\
Q_2 \,\, \mbox{M2:}\,\,\, & \_ & \_ & 3 & 4 & \_ & \_ & \_ & \, \\
Q_3 \,\, \mbox{M2:}\,\,\, & \_ & \_ & \_ & \_ & 5 & 6 & \_ & \, \\
q_1 \,\, \mbox{m5:}\,\,\, & \_ & \_ & 3 & 4 & 5 & 6 & \psi & \, \\
q_2 \,\, \mbox{m5:}\,\,\, & 1 & 2 & \_ & \_ & 5 & 6 & \psi & \, \\
q_3 \,\, \mbox{m5:}\,\,\, & 1 & 2 & 3 & 4 & \_ & \_ & \psi & \,.
\end{array}
\label{intersection}
\ee
We will denote by $z^i$ the coordinates along the 123456-directions,
which we take to span a six-torus. The three M5-branes wrap a common
circular direction, parametrized by $\psi$, in the four-dimensional
space transverse to the three M2-branes. Since this circle is
contractible, the M5-branes do not carry conserved charges but are instead
characterized, as we will see, by their dipoles $q_i$. The M2-branes
do carry conserved charges $Q_i$.

The $D=11$ supergravity solution describing this system takes the
form\footnote{
The action of $D=11$ supergravity is given in equation
\reef{eqn:11daction}.}
\ba
ds_{\it 11}^2 &=&  ds^2_{\it 5} + X^1 \left( dz_1^2 +
dz_2^2 \right) + X^2 \left( dz_3^2 + dz_4^2 \right) +
X^3 \left( dz_5^2 + dz_6^2 \right),\nonumber \\
\cala &=& A^1 \wedge dz_1 \wedge dz_2 + A^2 \wedge dz_3 \wedge dz_4 +
A^3 \wedge dz_5 \wedge dz_6 \,,
\label{11sol}
\ea
where $\cala$ is the three-form potential with four-form field
strength $\calf=d\cala$. The solution is specified by a metric
$ds^2_{\it 5}$, three scalars $X^i$, and three one-forms
$A^i$, with field strengths $F^i =dA^i$, which are defined on a five-dimensional
spacetime by
\ba
\label{eqn:5dsol}
ds_{\it 5}^2 &=& -(H_1 H_2 H_3)^{-2/3} (dt + \omega)^2 +
(H_1 H_2 H_3)^{1/3} d{\bf x}_{\it 4}^2,\nonumber \\
A^i &=&  H_i^{-1}(dt +\omega) -
\frac{q_i}{2} \left[(1+y) d\psi + (1+x) d\phi \right], \label{5sol} \\
X^i &=& H_i^{-1} (H_1 H_2 H_3)^{1/3} \,, \nonumber
\ea
where
\be
\label{eqn:flatspace}
d{\bf x}_{\it 4}^2 = \frac{R^2}{(x-y)^2} \left[ \frac{dy^2}{y^2-1} +
(y^2-1)d\psi^2 +\frac{dx^2}{1-x^2}+(1-x^2)d\phi^2 \right] \label{base}
\,,
\ee
\ba
H_1 &=& 1 + \frac{Q_1 - q_2 q_3 }{2R^2} (x-y) -
\frac{q_2 q_3}{4 R^2} (x^2 - y^2),\nn
H_2 &=& 1 + \frac{Q_2 - q_3 q_1 }{2R^2} (x-y) -
\frac{q_3 q_1}{4 R^2} (x^2 - y^2),\\
H_3 &=& 1 + \frac{Q_3 - q_1 q_2 }{2R^2} (x-y) -
\frac{q_1 q_2}{4 R^2} (x^2 - y^2),\nonumber
\label{eqn:Hi}
\ea
and $\omega = \omega_\phi d\phi + \omega_\psi d\psi$ with
\ba
 \omega_\phi &=& - \frac{1}{8 R^2} (1-x^2) \left[ q_1 Q_1 + q_2 Q_2 + q_3 Q_3 -
   q_1 q_2 q_3 \left( 3 + x + y
   \right) \right], \\
 \omega_\psi &=& \frac{1}{2} (q_1 + q_2 + q_3) (1+y)  - \frac{1}{8R^2}
(y^2-1) \left[
q_1 Q_1 + q_2 Q_2 + q_3 Q_3 - q_1 q_2 q_3 \left( 3 + x + y \right)
\right].\nonumber
\label{eqn:omegas}\ea
Note that the six-torus in \eqn{11sol} has constant
volume, since
\be
\label{eqn:constr}
 X^1 X^2 X^3 = 1.
\ee
This constraint implies that the five-dimensional metric $ds_{\it 5}^2$ is the
same as the Einstein-frame metric arising from reduction of the
above solution on $T^6$.
Note as well that, although the functions
$H_i$ are not harmonic, they appear in the metric \eqn{11sol} as
would be expected on the basis of the `harmonic superposition rule'
for the three M2-branes.

\begin{figure}[!t]
\begin{picture}(0,0)(0,0)
\footnotesize{
\put(13,34){$x=-1$}
\put(49,34){$x=+1$}
\put(33,40){$x$}
\put(59,64){$\psi$}
\put(83,63){$y$}
\put(53,6){$y=-1$}
\put(38,-3){$x=\mathrm{const}$}
}
\end{picture}
\centering{\epsfxsize=11cm\epsfbox{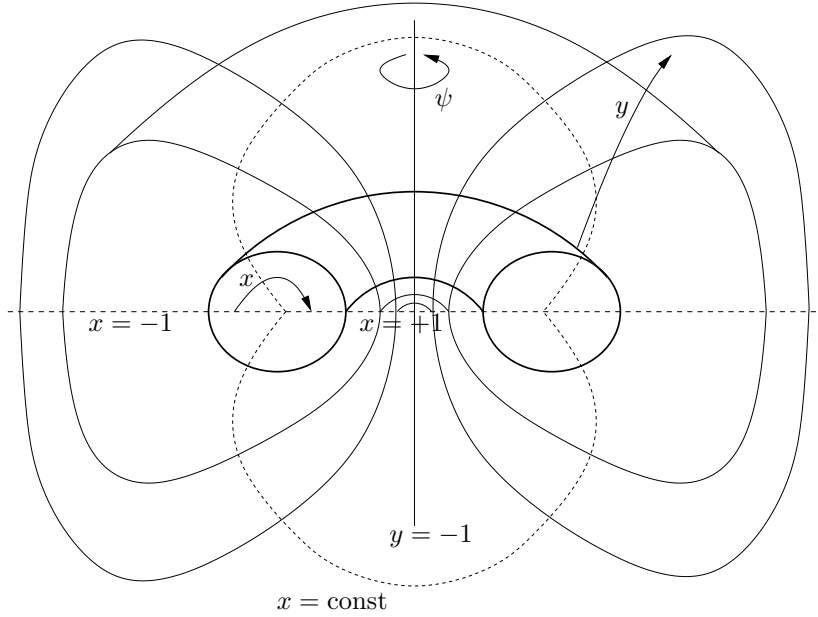}}
\vskip.5cm
\caption{\small Coordinate system for black ring metrics (from
\cite{ER2,RE}). The diagram sketches a section at constant $t$ and
$\phi$. Surfaces of constant $y$ are ring-shaped, while $x$ is a
polar coordinate on the $S^2$ (roughly $x\sim\cos\theta$). $x=\pm 1$ and
$y=-1$ are fixed-point sets (\ie axes) of $\partial_\phi$ and
$\partial_\psi$, respectively. Asymptotic infinity lies at $x=y=-1$.
}
\label{fig:coords}
\end{figure}

The metric $d{\bf x}_{\it 4}^2$ (which we shall sometimes refer to as
the ``base space") is just the flat metric on $\bbe{4}$
written in `ring coordinates' \cite{ER2,ER,HE,EE,RE}. These foliate
$\bbe{4}$ by surfaces of constant $y$ with topology $S^1\times S^2$,
which are equipotential surfaces of the field created by a ring-like
source. They are illustrated in fig.~\ref{fig:coords}.
The coordinates take values in the ranges $-1\leq x\leq 1$ and
$-\infty<y\leq -1$; $\phi,\psi$ are polar angles in two orthogonal
planes in $\bbe{4}$ and have period $2\pi$. Asymptotic infinity lies
at $x\to y\to -1$. Note that the apparent singularities at
$y=-1$ and $x = \pm 1$ are merely coordinate singularities, and that
$(x,\phi)$ parametrize (topologically) a two-sphere. The locus
$y=-\infty$ in the four-dimensional geometry \reef{eqn:flatspace} is a
circle of radius $R>0$ parametrized by $\psi$. We will show that in
the full geometry \reef{eqn:5dsol} this circle is blown up into a
finite-area, regular horizon. Note also that the function $x-y$ is
harmonic in \reef{eqn:flatspace}, with Dirac-delta sources on the
circle at $y=-\infty$.

The angular momentum one-form $\omega$ is globally well defined,
since $\omega_\phi(x=\pm 1)=\omega_\psi(y=-1)=0$, \ie there are no
Dirac-Misner strings. If these had been present then removing them
would have required a periodic identification of the time
coordinate, rendering the solution unphysical \cite{EE}. In
contrast, the potentials $A^i$ are not globally well defined since
there are Dirac strings at $x=+1$ (but not at $x=-1$ or $y=-1$).
This poses no problem, however, because their gauge-invariant field
strengths are well defined.

As mentioned above, $Q_i$ and $q_i$ are constants that measure the
charges and the dipole moments of the configuration. We assume that
\beq \label{Qqq}
Q_1\geq q_2 q_3\,,\quad Q_2\geq q_1 q_3\,,\quad Q_3\geq q_1 q_2\,,
\eeq
so that $H_i\geq 0$ (this assumption will be justified below). For later
convenience, we define
\be
\label{eqn:adef}
\mathcal{Q}_1 = Q_1 - q_2 q_3, \qquad \mathcal{Q}_2 = Q_2 - q_3 q_1,
\qquad \mathcal{Q}_3 = Q_3 -
 q_1 q_2\,,
\ee
which obviously satisfy $0\leq \mathcal{Q}_i\leq Q_i$, and
\be
q \equiv \left( q_1 q_2 q_3 \right)^{1/3}.
\ee
Integer powers of $q$ such as $q^2$ and $q^3$ should not be confused
with individual dipole moments, which we always label by a
subindex, \ie as $q_2$ and $q_3$.

It is shown in appendices \ref{app:minsugra} and \ref{app:U1N} that
the fields \eqn{11sol}, \eqn{eqn:5dsol} provide a supersymmetric
solution of $D=11$ supergravity. This is done as follows. First $D=11$
supergravity is reduced on $T^6$ using the ansatz \eqn{11sol} and the
constraint \eqn{eqn:constr}. This gives a ${\cal N}=1$
$D=5$ supergravity theory with gauge group $U(1)^3$ consisting
of $D=5$ minimal supergravity coupled to two
$U(1)$ vector multiplets. The bosonic fields of this theory are the metric, the
three abelian gauge fields $A^i$ and the three scalars $X^i$, which obey the
constraint \eqn{eqn:constr}.  This is a special case of a more general
$U(1)^N$ theory obtained by coupling minimal supergravity to $N-1$
vector multiplets. A general form for supersymmetric solutions of the
latter theory was obtained in \cite{gutowski:04a,gutowski:04b},
generalizing the results of \cite{harveyetal} for the minimal
theory. Using these results, it is a simple task to extend our
construction of the supersymmetric black ring solution from the
minimal theory to this more general theory. The general $N$-charge
supersymmetric black ring solution is given in appendix
\ref{app:U1N}. For the special case of the $U(1)^3$ theory, it reduces
to the solution \eqn{eqn:5dsol}. The analysis of
\cite{gutowski:04a,gutowski:04b} reveals that all supersymmetric
solutions of the $D=5$ theory preserve either four or eight
supersymmetries, and a complete list of the latter was given in
\cite{gutowski:04a}. It follows that our solution preserves four
supersymmetries and hence gives a $1/8$ BPS solution of $D=11$
supergravity.

In the special case of three equal charges $Q_i = Q$ and three equal
dipoles $q_i = q$, the $D=5$ solution \eqn{eqn:5dsol} reduces to the
supersymmetric black ring solution of minimal supergravity constructed
in \cite{EEMR}. We shall show that the general seven-parameter solution
\eqn{eqn:5dsol} also describes supersymmetric black rings.

\setcounter{equation}{0}
\section{Physical Properties}
\label{sec:phys}

In this section we shall compute the physical quantities that
characterize the black supertube solution, determine the necessary and
sufficient conditions to avoid causal pathologies and demonstrate that
it has a regular horizon. We first introduce some new coordinate systems
that are useful for different aspects of the analysis.

\subsection{Coordinate systems}
\label{subsec:coords}

The coordinates employed in the previous section display the
solution in a form that involves simple functions of $x$ and $y$ and
indeed provide the easiest way to derive it. To obtain the charges
measured at infinity, however, it is convenient to introduce
coordinates in which the asymptotic flatness of the solution becomes
manifest. Specifically, we change $(x,y) \ra (\rho,\Theta)$ through
\be
  \rho \sin{\Theta} = \frac{R\sqrt{y^2-1}}{x-y} \, ,~~~~
  \rho \cos{\Theta} = \frac{R\sqrt{1-x^2}}{x-y} \, ,
  \label{rhoTheta}
\ee
with $0\leq \rho<\infty$, $0\leq \Theta\leq \pi/2$. Define also
\beq
\Sigma \equiv \frac{2R^2}{x-y}=
\sqrt{(\rho^2-R^2)^2 + 4R^2 \rho^2 \cos^2{\Theta}}\,.
\label{Sigma1}\eeq
\begin{figure}[t]%
\begin{picture}(0,0)(0,0)%
\footnotesize{
\put(28,51){$\Theta=0$}%
\put(62,24){$\Theta=\pi/2$}%
\put(-12,24){$\Theta=\pi/2$}%
\put(14,26){$\rho=R$}%
}
\end{picture}%
\centering{\epsfxsize=6cm\epsfbox{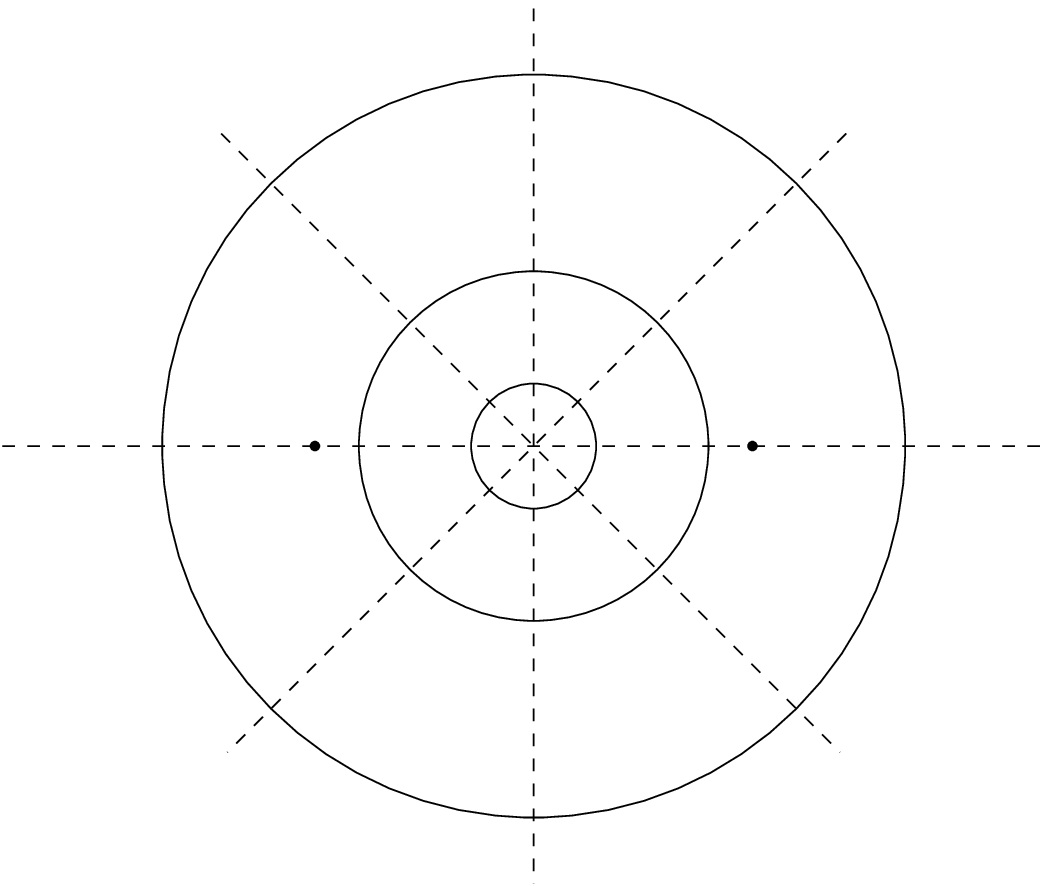}}
\vskip.5cm
\caption{\small Coordinates $(\rho,\Theta)$, in a section at constant
$t$, $\phi$, $\psi$ (the four quadrants are obtained by including also
constant $\phi+\pi$ and $\psi+\pi$). Solid lines are surfaces of constant
$\rho$, dashed lines are at constant $\Theta$. The ring lies at
$\rho=R$, $\Theta=\pi/2$.}%
\label{fig:rhoTheta}%
\end{figure}%
In these coordinates the flat base space metric is
\beqa
  d{\bf x}_4^2 = d\rho^2
  + \rho^2 (d\Theta^2 + \sin^2{\Theta}d\psi^2+\cos^2{\Theta}d\phi^2)
\, ,
\label{rhoThetabase}
\eeqa
The functions entering the solution are
\beqa
  H_1 &=& 1+ \frac{Q_1-q_2 q_3}{\Sigma}
  +q_2 q_3\frac{\rho^2}{\Sigma^2} \, , \nn
  H_2 &=& 1+ \frac{Q_2-q_3 q_1}{\Sigma}
  +q_3 q_1\frac{\rho^2}{\Sigma^2} \, ,\\
  H_3 &=& 1+ \frac{Q_3-q_1 q_2}{\Sigma}
  +q_1 q_2\frac{\rho^2}{\Sigma^2} \nonumber\, ,
\eeqa
\beqa
  \omega_\phi &=& - \frac{\rho^2 \cos^2{\Theta}}{2 \Sigma^2}
    \left[ q_1 Q_1 + q_2 Q_2 + q_3 Q_3
      - q^3 \left( 3- \frac{2\rho^2}{\Sigma}\right)\right]
  \, , \\
  \omega_\psi &=&
    -(q_1+q_2+q_3)\frac{2 R^2 \rho^2 \sin^2{\Theta}}
          {\Sigma (\rho^2+R^2+\Sigma)}
    - \frac{\rho^2 \sin^2{\Theta}}{2 \Sigma^2}
    \left[ q_1 Q_1 + q_2 Q_2 + q_3 Q_3
      - q^3 \left( 3- \frac{2\rho^2}{\Sigma}\right)\right]
  \, ,\nonumber
\eeqa
\beq
A^i =  H_i^{-1}(dt +\omega) +
\frac{q_i}{2\Sigma} \left[(\rho^2+R^2-\Sigma) d\psi +
(\rho^2-R^2-\Sigma)d\phi \right].
\eeq
Note that $\Sigma^{-1}$ is a harmonic function in
\reef{rhoThetabase} with Dirac-delta sources on a ring at $\rho=R$,
$\Theta=\pi/2$ (see figure \ref{fig:rhoTheta}). As $\rho \ra \infty$
the five-dimensional metric \eqn{eqn:5dsol} is manifestly
asymptotically flat, and the eleven-dimensional metric \eqn{11sol}
is asymptotically flat in the directions transverse to all of the
M2-branes.

\begin{figure}[t]
\begin{picture}(0,0)(0,0)
\footnotesize{
\put(28,51){$\theta=0$}
\put(62,24){$\theta=\pi/2$}
\put(-12,24){$\theta=\pi/2$}
\put(28,25){$r=0$}
\put(45,39){$r=\mathrm{const}$}
}
\end{picture}
\centering{\epsfxsize=6cm\epsfbox{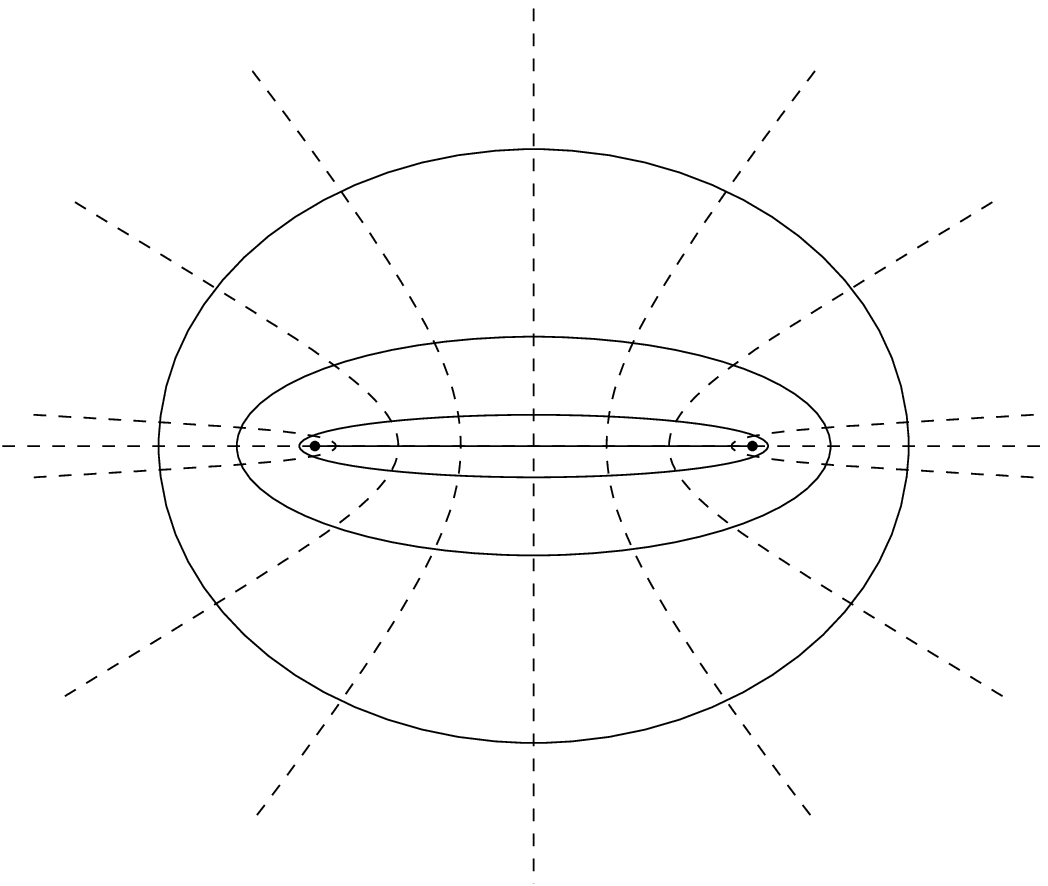}}
\vskip.5cm
\caption{\small Coordinates $(r,\theta)$, in a section at constant $t$,
$\phi$, $\psi$ (and $\phi+\pi$, $\psi+\pi$). Solid lines are surfaces
of constant $r$,
dashed lines are at constant $\theta$. The axis of $\phi$ consists of
the segments $r=0$ and $\theta=\pi/2$.
}
\label{fig:rtheta}
\end{figure}

This coordinate system foliates $\bbe{4}$ in a familiar manner, but
is quite unwieldy for studying the structure of the solution near
the ring. There is yet a third system of coordinates that proves
useful for later applications, in particular for describing the
decoupling limit in section
\ref{subsec:decoupling}. These coordinates are defined by changing
$(x,y)\to (r,\theta)$ through
\beq
r^2=R^2\frac{1-x}{x-y}\,,\qquad \cos^2\theta=\frac{1+x}{x-y}\,,
\label{rtheta}
\eeq
where $0 \leq r < \infty, 0 \leq \theta \leq \pi/2$. The flat base
space metric is
\beq
d{\bf x}_4^2 = \Sigma\left(\frac{dr^2}{r^2+R^2}+d\theta^2\right)
  + (r^2+R^2) \sin^2{\theta}d\psi^2+r^2\cos^2{\theta}d\phi^2
\, ,
\label{rthetabase}\eeq
where now the function $\Sigma$ defined in \reef{Sigma1} takes the form
\beq
\Sigma=r^2+R^2\cos^2\theta\,,
\eeq
and does not involve any surds. Surfaces at constant $r$ are
topologically $S^3$'s that enclose the ring. The inner disk of the
ring, $\{x=+1\}$, corresponds now to $\{r=0, 0< \theta< \pi/2\}$,
and the outer annulus, $\{x=-1\}$, is $\{0<r<\infty, \theta=\pi/2\}$
(see figure \ref{fig:rtheta}). The horizon of the ring lies at
$r=0$, $\theta=\pi/2$. The full solution becomes manifestly flat as
$r\to\infty$, where $r$ and $\theta$ come to coincide with the previous
$\rho$ and $\Theta$.
The functions defining the solution are now
\beq
H_1=1+\frac{Q_1}{\Sigma}-\frac{q_2q_3R^2\cos 2\theta}{\Sigma^2}\,,
\eeq
with the obvious permutations of (123) giving $H_2$ and $H_3$, and
\beqa
\label{omegarth}
\omega_\phi &=& - \frac{r^2 \cos^2\theta}{2\Sigma^2}
    \left[q_1 Q_1 + q_2
Q_2 + q_3 Q_3 - q^3\left(1+\frac{2R^2\cos 2\theta}{\Sigma}\right)\right]
  \, , \\
  \omega_\psi &=&
    -(q_1+q_2+q_3)\frac{R^2\sin^2\theta}{\Sigma}
    - \frac{(r^2+R^2) \sin^2{\theta}}{2\Sigma^2}
    \left[q_1 Q_1 + q_2
Q_2 + q_3 Q_3 - q^3\left(1+\frac{2R^2\cos 2\theta}{\Sigma}\right)\right]
  \, .\nonumber\eeqa
For convenience, we also give the gauge potentials
\beq
\label{Arth}
A^i = H_i^{-1}(dt +\omega) +
 \frac{q_i R^2}{\Sigma} (\sin^2\theta d\psi -\cos^2\theta d\phi )\,.
\eeq

\subsection{Physical parameters}

If we assume that the $z_i$ directions are all compact with length
$2\pi\ell$, then the five-dimensional Newton's constant $G_5$ is
related to the 11D coupling constant $\kappa$ through $\kappa^2=8\pi
G_5 (2\pi\ell)^6$. The mass and angular momenta in five dimensions
can be read off from the asymptotic form of the above metric,
\ba
M &=& \frac{\pi}{4G_5} (Q_1 + Q_2
+ Q_3), \nonumber \\
\label{MQJ}
J_{\phi} &=& \frac{\pi}{8G_5} \left(q_1 Q_1 + q_2
Q_2 + q_3 Q_3 - q_1 q_2 q_3 \right),\\
\qquad J_{\psi} &=&
\frac{\pi}{8G_5} \left( 2R^2 (q_1 +q_2 + q_3) + q_1 Q_1 + q_2 Q_2 + q_3
Q_3 - q_1 q_2 q_3 \right).\nonumber
\ea
The M2-brane charges carried by the solution are given by
\be
\mathbf{Q}_i \equiv
\frac{(2\pi\ell)^{2}}{2\kappa^2} \int_{S^3 \times T^4}
\star_{11} \calf =
\frac{1}{16\pi G_5} \int_{S^3} (X^i)^{-2} \star_5 F^i  = \frac{\pi}{4G_5} Q_i,
\ee
where $\star_{11}$ and $\star_5$ are the eleven- and
five-dimensional Hodge dual operators with respect to the metrics
\eqn{11sol} and \eqn{eqn:5dsol}, respectively. The $S^3$ is the
sphere at infinity in the five-dimensional spacetime, and $T^4$
denotes the 3456, 1256 and 1234 four-torus for $i=1,2,3$,
respectively. The solution saturates the BPS bound
\be
M=\mathbf{Q}_1 + \mathbf{Q}_2 + \mathbf{Q}_3 \,.
\ee
As we have explained, the M5-branes do not carry any net charges.
However, their presence can be characterized by appropriate fluxes,
to which we will refer as `dipole charges', through surfaces that
encircle the ring once, namely by
\be
\mathbf{D}_i \equiv \frac{(2\pi\ell)^{4}}{2\kappa^2} \int_{S^2 \times
T^2} \calf =
\frac{1}{16 \pi G_5} \int_{S^2} F^i = \frac{q_i}{8G_5}\,,
\ee
where the $S^2$ is a surface of constant $t$, $y$ and $\psi$ in the
metric \reef{11sol}, and $T^2$ is a two-torus in the 12-, 34- and
56-directions for $i=1,2,3$, respectively. In computing these
integrals it is useful to observe that the first summand in $A^i$
does not contribute, since, because $\omega$ is globally well
defined, it leads to a total derivative.

These dipole constituents generate a dipole field component of $\cal
F$ near infinity. Asymptotically, the magnetic components of the
three-form potential are given by
\bea
A^i_\phi&\to& -\left( \frac{4G_5J_\phi}{\pi} +q_i R^2 \right)
\frac{\cos^2\Theta}{\rho^2} \,, \label{aphi} \\
A^i_\psi&\to&-\left( \frac{4G_5J_\psi}{\pi} -q_i R^2 \right)
\frac{\sin^2\Theta}{\rho^2} \,.
\label{apsi}
\eea
The presence of non-zero $\phi$-components is easily understood.
Consider for example $A^1_\phi$ (the interpretation for $A^2_\phi$
and $A^3_\phi$ is analogous). This corresponds to a non-zero
$\cala_{12\phi}$ that, upon Hodge dualization, leads to a non-zero component
${\tilde{\cala}}_{03456\psi}$, as expected for the potential sourced by M5-branes
along the $3456\psi$-directions, as in the array \eqn{intersection}.
The magnitude of this dipole moment is set by the coefficient in
brackets in \eqn{aphi}. The second contribution, $q_1 R^2$, is
exactly as would be expected for a one-dipole supertube source
\cite{EMT,MNT}. The interpretation of the contribution proportional to
$J_\phi$ is more subtle, and its origin can presumably be understood
in the same way as that of $J_\phi$ itself, which will be discussed
in sec. \ref{sec:sugravsworld}.

Hodge dualization of the $A^i_\psi$ components would seemingly
suggest the presence of M5-brane sources that wrap the
$\phi$-direction. However, examination of the details of
the supergravity solution reveals that there are no such sources.
Instead, the correct interpretation of these components is that they
are sourced by the M2-branes in the presence of the M5-branes.
Consider, for example, a two-charge/one-dipole supertube consisting
of the M2-branes along the 12- and 34-directions and the M5-brane
along the $1234\psi$-directions. The M2-branes can be represented by
fluxes $H_{012}$ and $H_{034}$ of the M5-brane worldvolume
three-form. These couple minimally to, and hence act as sources of,
the $\mathcal{A}_{12\psi}$ and $\mathcal{A}_{34\psi}$ components of
the supergravity potential through the Wess-Zumino term of the
M5-brane action, $S_{WZ} \sim \int \mathcal{A}_3 \wedge H_3$.

We can infer from the expressions \reef{aphi}-\reef{apsi}
that the gyromagnetic ratio of the supersymmetric ring is
$g=3$, as for the BMPV black hole \cite{carh}.

In the quantum theory the charges will be integer-quantized,
with \cite{KT}
\be
N_i=\left(\frac{\pi}{4G_5}\right)^{2/3} Q_i\,,
\label{quantN}\ee
\be
n_i=
\left(\frac{\pi}{4G_5}\right)^{1/3} q_i\,,
\label{quantn}\ee
corresponding to the numbers of M2- and M5-branes in the system,
respectively.

\subsection{Causal structure and horizon geometry}
\label{subsec:nocccs}

The results of \cite{harveyetal} reveal that many rotating
supersymmetric solutions exhibit closed causal curves
(CCCs). In this section, we shall derive a simple criterion for the absence
of such pathologies in a general five-dimensional supersymmetric solution
and then examine when our solution \eqn{eqn:5dsol} satisfies this criterion.

Any supersymmetric solution of $D=5$ supergravity theory admits a
non-spacelike Killing vector field $V$ \cite{gibbons:93, gutowski:04a},
which defines a preferred time orientation. In a region where $V$ is
timelike, the metric can be written as (see appendix A for details)
\be
ds^2 = -f^2 (dt + \omega)^2 + f^{-1} h_{mn} dx^m dx^n
\ee
where $V = \partial/\partial t$ and $h_{mn}$ is a
Riemannian metric on a four-dimensional space with coordinates
$x^m$. The metric $h_{mn}$, scalar $f$ and 1-form $\omega \equiv
\omega_m dx^m$ are all independent of $t$. For our solution, $f^{-1} =
(H_1 H_2 H_3)^{1/3}$, $x^m=\{\psi, y, \phi, x\}$ and $h_{mn}$ is
flat.

Consider a smooth, future-directed, causal curve in such a region. Let $U$
denote the tangent to the curve and $\lambda$ a parameter along the
curve. Then, using a dot to denote a derivative with respect to
$\lambda$, we have
\be
\label{eqn:futuredir}
 0 \le - V \cdot U = f^2 \left( \dot{t} + \omega_m \dot{x}^m \right),
\ee
because the curve is future-directed. Furthermore,
\be
\label{eqn:Usq}
 f^2 \dot{t} \left( \dot{t} + 2 \omega_m \dot{x}^m \right) = -U^2 +
g_{mn} \dot{x}^m \dot{x}^n,
\ee
where
\be
 g_{mn} \equiv f^{-1} h_{mn} - f^2 \omega_m \omega_n.
\ee
Let us now assume that $g_{mn}$ is positive-definite. Then the
right-hand-side of equation \reef{eqn:Usq} is non-negative because
$U^2 \le 0$. We shall show that this implies $\dot{t} >0$. Consider
first the special case in which the RHS of \reef{eqn:Usq} vanishes.
This implies that $\dot{x}^m = 0$. Equation
\reef{eqn:futuredir} then gives $\dot{t}>0$ (we cannot have $\dot{t}=0$
as that would imply $U=0$). Now consider the general case in which the
RHS of \reef{eqn:Usq} is positive. Then either (i) $\dot{t}>0$ and
$\dot{t} + 2 \omega_m \dot{x}^m>0$ or (ii) $\dot{t}<0$ and $\dot{t}
+2 \omega_m \dot{x}^m <0$. However, it is easy to see that (ii) is
inconsistent with \reef{eqn:futuredir}. Hence we must have (i) so
$\dot{t} >0$. Therefore $t$ must increase along any causal curve, so
if $t$ is globally defined then such a curve cannot intersect itself.

In summary, the condition that $g_{mn}$ be positive-definite is sufficient
to ensure that there are no closed causal curves contained entirely
within a region in which $V$ is timelike and $t$ is globally defined.

For our solution, the coordinates $(t,x,y,\phi,\psi)$ cover such a
region. Hence to show that our solution has no CCCs at finite $y$, it is
sufficient to show that $g_{mn}$ is positive-definite for
$-\infty<y<-1$. This reduces to
showing that $g_{ij}$ is positive-definite, where $i,j$ are the $\phi,\psi$
directions. In Appendix \ref{app:noctcs} we show that a necessary and sufficient
condition for $g_{ij}$ to be positive-definite is\footnote{
We assume $q_i>0$ so the inequality \eqn{eqn:noctcs} requires $\mathcal{Q}_i$ to
lie in the region interior to one of the sheets of a two-sheeted
hyperboloid in $\mathbb{R}^3$. One sheet lies entirely in the positive octant of
$\mathbb{R}^3$ (i.e. $\mathcal{Q}_i>0$) and the other entirely in the negative octant. We
have assumed that we are dealing with the positive octant. This
justifies our earlier restriction \eqn{Qqq}.}
\be
  2 \sum_{i < j} \mathcal{Q}_i q_i \mathcal{Q}_j q_j - \sum_i
  \mathcal{Q}_i^2 q_i^2 \ge 4 R^2 q^3 \sum_i q_i,
\label{eqn:noctcs}
\ee
where we use the $\mathcal{Q}_i$ defined in \reef{eqn:adef}. It follows from the
above argument that our solution is free of CCCs at finite $y$ if
the inequality \reef{eqn:noctcs} is satisfied. This might be regarded as
providing an upper bound on the radius $R$ for a given set of charges
$Q_i$ and $q_i$. However, as we shall see, $R$ is not the physical radius of the ring.
An equivalent expression that involves only physical quantities is
\beq
4q_1q_2\left(\mathcal{Q}_1\mathcal{Q}_2-q_3\frac{4G_5(J_\psi-J_\phi)}{\pi}\right)\geq
(\mathcal{Q}_1q_1+\mathcal{Q}_2q_2-
\mathcal{Q}_3q_3)^2\,.
\label{eqn:noctcsb}
\eeq
This expression yields now an upper bound on $J_\psi-J_\phi$ for given
charges. Other equivalent forms are obtained by permutations of (123).

As $y \rightarrow - \infty$ we find
\be
 g_{\psi \psi} = L^2 + \frac{q^2}{4} (1-x^2) + {\cal O}\left(\frac{1}{y} \right),
\ee
where
\ba
 L &\equiv&  \frac{1}{2 q^2} \left[2 \sum_{i < j} \mathcal{Q}_i q_i
 \mathcal{Q}_j q_j - \sum_i \mathcal{Q}_i^2 q_i^2 - 4R^2 q^3 \sum_i q_i \right]^{1/2} \nn
 &=& \frac{1}{q^2}
\left[q_1q_2\left(\mathcal{Q}_1\mathcal{Q}_2-q_3\frac{4G_5(J_\psi-
J_\phi)}{\pi}\right)-\frac{1}{4}
(\mathcal{Q}_1q_1+\mathcal{Q}_2q_2-
\mathcal{Q}_3q_3)^2 \right]^{1/2},
\label{bigL}
\ea
which is real and non-negative as a consequence of \reef{eqn:noctcs}. If
\eqn{eqn:noctcs} were violated then $\partial/\partial \psi$ would
become timelike as $y \rightarrow -\infty$ in a neighbourhood of $x = \pm 1$ so some
orbits of $\partial/\partial\psi$ would be closed timelike curves.

In \cite{EEMR} we showed that the BPS ring solution of minimal
supergravity can be analytically extended through an event horizon at
$y = -\infty$ when the inequality \reef{eqn:noctcs} is strict (\ie when $L>0$). The
same is true of the general solution presented above. The method of
extending the solution is the same as in the minimal theory --- the
details are presented in Appendix \ref{app:through}.
There we show that the geometry of a spacelike section
of the horizon is the product of a circle of radius $L$ and a round
2-sphere of radius $q/2$,
\be
ds_{H}^2 = L^2 d{\psi'}^2 + \frac{q^2}{4}
\left(d\bar{\theta}^2+\sin^2\bar{\theta} d\chi^2\right),
\ee
The ring circle is parametrized by the coordinate $\psi'$, which is a
good coordinate on the horizon, while $\psi$ itself is not. These
coordinates differ by a function of $y$ (see \eqn{EFcoords} for details)
and hence have the same period $2\pi$. The $S^2$ coordinates are
$\chi=\phi-\psi$ and $\cos\bar\theta=x$, which are well-behaved at the
horizon.
The horizon area is
\beq
  \mathcal{A}_H = 2 \pi^2 \, L \,q^2\,.
\label{ringarea}\eeq
Observe that the proper circumferential length of the ring horizon is
$2\pi L$, not $2\pi R$, and can be arbitrarily large or small for fixed $R$.
In a form more symmetric in the three constituents,
\beq
L=\frac{1}{q^2}\left[Q_1Q_2Q_3-\mathcal{Q}_1\mathcal{Q}_2\mathcal{Q}_3
-\left(\frac{4G_5J_\phi}{\pi}\right)^{2}
-q^3\frac{4G_5(J_\psi-J_\phi)}{\pi}\right]^{1/2}\,.
\label{bigLb}
\eeq
When $L=0$, it is shown in appendix \ref{app:nullorb} that the solution has a null orbifold singularity instead of a regular event horizon.

Finally we should mention that the restriction \eqn{eqn:noctcs}
guarantees only that CCCs are absent in the region exterior to the event
horizon of the black ring. There will certainly be CCCs present behind
the event horizon.

\setcounter{equation}{0}
\section{The double helix: D1-D5-P black supertube}
\label{sec:d1d5p}
The solution of eleven-dimensional supergravity given in section
\ref{sec:gensol} can be Kaluza-Klein reduced to a solution of type IIA
supergravity and then dualized to a type IIB solution with net charges
D1, D5 and momentum (P). We study here the properties of
this black supertube solution.

\subsection{The IIB solution}

Perform a KK reduction of \reef{11sol} along $z_6$, and T-dualize on
$z_5$, $z_4$, $z_3$ (using \cite{bergshoeff:95}) to get a IIB
supergravity solution. The solution has D1-D5-P charges and D1, D5 and
Kaluza-Klein monopole (kkm) dipoles.
The D1-D5-P supergravity solution describes a three-charge
black supertube.
In string theory, a D1-D5-P supertube is actually a
double D1-D5 helix that carries momentum in the direction parallel to
its axis, along $z\equiv z_5$, and which coils around the direction of
the ring $\psi$. The D1 and D5 branes are bound to a tube made of KK
monopoles spanning the ring circle and $z_1,z_2,z_3,z_4$, with the
direction $z$ being the $U(1)$ fiber of the KK monopoles.
In array form
\be
\begin{array}{llccccccl}
Q_1  &\mbox{D5:} \,\, & z  & 1  & 2  & 3  & 4  & \_   & \, \\
Q_2  &\mbox{D1:} \,\, & z  & \_ & \_ & \_ & \_ & \_   & \, \\
Q_3  &\mbox{P:}  \,\, & z  & \_ & \_ & \_ & \_ & \_   & \, \\
q_1  &\mbox{d1:} \,\, & \_ & \_ & \_ & \_ & \_ & \psi & \, \\
q_2  &\mbox{d5:} \,\, & \_ & 1  & 2  & 3  & 4  & \psi & \, \\
q_3  &\mbox{kkm:}\,\, &(z) & 1  & 2  & 3  & 4  & \psi & \,.
\end{array}
\label{supertube}
\ee
Dualizing the supergravity solution as described above, we find that
the string frame metric of the D1-D5-P black supertube is
\ba
\label{eqn:2bmetric}
 ds^2 &=& - (X^3)^{1/2} ds_{\it 5}^2
          + (X^3)^{-3/2} \left(dz + A^3\right)^2
      + X^1 (X^3)^{1/2} d{\bf z}_4^2 \nonumber \\
&=& - \frac{1}{H_3 \sqrt{H_1 H_2}}(dt + \omega)^2
          + \frac{H_3}{\sqrt{H_1 H_2}}\left(dz + A^3\right)^2
          + \sqrt{H_1 H_2} \, d{\bf x}_4^2
      +\sqrt{\frac{H_2}{H_1}} \, d{\bf z}_4^2 \, ,
\ea
where $ds_5^2$, $X^i$, and $A^i$ are given in \reef{5sol}.\footnote{
Any supersymmetric solution of IIB supergravity must admit a globally
defined null Killing vector field \cite{jones:04}. For this
solution it is easy to see that $\partial/\partial t$ is globally
null.} The other non-vanishing fields are the dilaton and RR 3-form
field strength:
\be
  e^{2\Phi} = \frac{H_2}{H_1} \, \qquad
F^{(3)} = \left( X^1 \right)^{-2} \star_5 F^1 + F^2 \wedge \left(
dz+A^3 \right).
\label{2bfields}\ee
The Bianchi identity and equation of motion of $F^{(3)}$ are satisfied
as a consequence of the Bianchi identities and equations of motion of
the $D=5$ gauge fields $F^1$ and $F^2$.

This solution can be S-dualized to give a purely NS-NS solution of type II supergravity, and it then describes an
F1-NS5-P supertube. A trivial T-duality along any of the flat directions ${\bf z}_4$ of the NS5 brane maps the
F1-NS5-P supertube to a solution of IIA supergravity, which when uplifted to eleven dimensions along a direction
$\tilde{z}$ provides an embedding in D=11 supergravity different than the one in \reef{11sol}. This new embedding
describes M2 and M5 branes that intersect over a helical string, and which are bound to a tube of KK monopoles.
Reducing it along $z$ yields a supertube with D0-F1-D4 charges bound to D2-D6-NS5 tubular branes. T-dualizing this
along $\tilde{z}$ gives back a D1-D5-P supertube with the charges $Q_i$ shuffled compared to the first
configuration \reef{supertube}-\reef{eqn:2bmetric}.

Due to its particular relevance to the microscopic CFT description of
black holes, in the following we will mostly focus on the D1-D5-P version of
the solution. We assume that the directions ${\bf z}_4$ are
compact with length\footnote{For simplicity we use the same letter $\ell$
to denote the compact radii in the IIB solution as in the $D=11$
solution, even if they are not invariant under the dualities that relate
them. We hope that this does not cause any confusion.}
 $2\pi\ell$, while the length along $z$ is $2\pi R_z$.
The numbers of D5 and D1 branes and momentum units are then
\be
\quad N_\mathrm{D5}=\frac{1}{g_s \ell_s^2}\, Q_1\,,\qquad
N_\mathrm{D1}=\frac{1}{g_s \ell_s^2}\left(\frac{\ell}{\ell_s}\right)^4\, Q_2\,,
\qquad
N_\mathrm{P}=\frac{1}{g_s^2
\ell_s^2}\left(\frac{R_z}{\ell_s}\right)^2
\left(\frac{\ell}{\ell_s}\right)^4\, Q_3\,,
\label{quantN15}
\ee
and the dipole components
\beq
n_\mathrm{D1}=
\frac{1}{g_s \ell_s}
\left(\frac{R_z}{\ell_s}\right)\left(\frac{\ell}{\ell_s}\right)^4\,
q_1\,,\qquad
n_\mathrm{D5}=
\frac{1}{g_s\ell_s}\left(\frac{R_z}{\ell_s}\right)\, q_2\,,
\label{quantn15}
\eeq
where $g_s$ and $\ell_s$ are the string coupling constant and string
length. The quantization condition on the KKM dipole will be rederived
below.

\subsection{Structure of the D1-D5-P supertube}
\label{subsec:d1d5struc}

\subsubsection{KK dipole quantization}

The solution \reef{eqn:2bmetric} possesses non-trivial structure along
the sixth direction $z$, so it is more appropriately viewed from a
six-dimensional perspective. The quantization of the KK dipole charge
follows then from purely
geometric considerations \cite{EE}. The metric $ds_\mathit{5}^2$
is clearly regular at $x=-1$ and $y=-1$. However, $A^3$
is not regular at $x=1$ unless we perform a gauge transformation. This
gauge transformation is a shift in the coordinate $z$:
\be
\label{eqn:ztransfm}
 z \rightarrow \hat{z} \equiv z - q_3 \phi
\ee
under which the dangerous terms transform as
\be
 dz - \frac{q_3}{2} (1+x) d\phi = d\hat{z} + \frac{q_3}{2} (1-x) d\phi,
\ee
which is now regular at $x=1$. However, $z$ parametrizes a compact
Kaluza-Klein direction, so $z \sim z + 2\pi R_z$. This implies
that the coordinate transformation \reef{eqn:ztransfm} is globally
well-defined only if
\be
 q_3 = n_\mathrm{KK} R_z,
\label{q3KK}\ee
for some positive integer $n_\mathrm{KK}$ (as we know $q_3>0$). Hence
the dipole charge $q_3$ is quantized in units of the radius of the KK
circle, as expected for a KK monopole charge. This is dual to the
quantization conditions \reef{quantn15}.

\subsubsection{Horizon geometry}

The $D=5$ no-CCC condition \eqn{eqn:noctcs} is sufficient to ensure that
the IIB solution is also free of naked CCCs because the extra terms in
the metric \eqn{eqn:2bmetric} are manifestly positive.\footnote{ We
shall not determine whether this condition is also {\it necessary} for
CCCs to be absent in the IIB solution since the $D=5$ description seems
to be more relevant (and more stringent) for the purpose of analyzing
causal anomalies. In the case of the BMPV solution it is known that CCCs
can be removed from the IIB solution by working in the universal
covering space, and only appear when the $z$ direction is compactified
\cite{carh}. This is not the case here. For instance, the $\psi\psi$
component of the IIB metric is not automatically positive near the
horizon without imposing some condition --- and \eqn{eqn:noctcs} is
sufficient for this.} Subject to the quantization condition \reef{q3KK},
the IIB solution is regular at finite $y$. As $y \rightarrow -\infty$,
the conformal factors multiplying the three terms in the first line of
\reef{eqn:2bmetric} remain finite and non-zero (since they just involve
powers of the $D=5$ scalar fields). We know that $ds_{\it 5}^2$ is
regular at $y=-\infty$ when $L>0$ so it remains only to show that
$dz+A^3$ is also regular there. The gauge transformation that achieves
this is described in appendix \ref{app:through}. It is then apparent
that $y=-\infty$ is an event horizon of the IIB solution.

Following appendix~\ref{app:through}, the geometry of a spatial slice
through the event horizon is (in string frame)
\bea
ds^2 &=&\frac{qL^2 }{\sqrt{q_1q_2}} d{\psi'}^2 +
\frac{q_3\sqrt{q_1q_2}}{4} \left( d\bar{\theta}^2 + \sin^2
\bar{\theta} d\chi^2 \right) \\
&&+
\frac{\sqrt{q_1q_2}}{q_3}\left[dz' - \frac{q_3}{2} (1+\cos \bar{\theta}) d\chi -
\frac{q_1\mathcal{Q}_1+q_2\mathcal{Q}_2-q_3\mathcal{Q}_3}{2q_1q_2} d\psi'
\right]^2 + \sqrt{\frac{q_1}{q_2}}\;d{\bf z}_4^2\,, \nonumber
\eea
where, recall, $q\equiv (q_1q_2q_3)^{1/3}$ and $\cos\bar{\theta}=x$. The coordinate $z'$ differs from $z$ only by a function of $y$, and hence has the same period $2\pi R_z$. The
coordinate transformation
\be
\label{eqn:zppdef}
z'' = z' -
\frac{q_1\mathcal{Q}_1+q_2\mathcal{Q}_2-q_3\mathcal{Q}_3}{2q_1q_2} \psi'
\ee
reveals that this is {\it locally} a product of $S^1$, parametrized by
$\psi'$, with a locally $S^3$ geometry parametrized by
$(z'',\bar{\theta},\chi)$ (and $T^4$). The locally $S^3$ part is only globally
$S^3$ in the special case $n_\mathrm{KK}=1$. For $n_\mathrm{KK}>1$ it
is a homogeneous lens space $S^3 /Z_{n_\mathrm{KK}}$. Note, however, that
the coordinate transformation \reef{eqn:zppdef} is not {\it globally}
well-defined unless
\be
q_1\mathcal{Q}_1+q_2\mathcal{Q}_2-q_3\mathcal{Q}_3= 2 q_1q_2m R_z,
\label{quantm}
\ee
with $m$ an integer. In this special case the horizon geometry is the
product $S^1 \times (S^3 /Z_{n_\mathrm{KK}})$. If this equation is not
satisfied then the horizon geometry is given by a regular non-product
metric on $S^1 \times (S^3 /Z_{n_\mathrm{KK}})$. \footnote{Formally,
this is the same as the oxidation of BMPV: see equation (6.8) of
\cite{gutowski:03} with $u \rightarrow \psi'$, $\psi' \rightarrow 2
z''/q_3$, $\phi \rightarrow \chi$. (Lens spaces do not arise from an
asymptotically flat BMPV black hole but they do arise from obvious
quotients of BMPV. They can also arise in the near-horizon geometry of
oxidized $D=4$ black holes \cite{cvetic:99}.)} To avoid confusion, we
emphasize that equation \eqn{quantm} does {\it not} have to be
satisfied in general but, when it is satisfied, the geometry of the
horizon factorizes. It is worth observing that in this case the no-CCC
bound \reef{eqn:noctcsb}, and also the expression for the entropy,
simplify considerably.

The near-horizon limit of the IIB solution is obtained by defining
$y=-R^2/(\epsilon L \tilde{r})$, $t=\tilde{t}/\epsilon$ and $\epsilon
\rightarrow 0$. In terms of the coordinates regular at the horizon
introduced in appendix \ref{app:through}, we take $\bar{r} = \epsilon L
\tilde{r} /R$, $v = \tilde{v} /\epsilon$. In this limit we obtain a
locally AdS$_3 \times S^3 \times T^4$ spacetime\footnote{This is the
string frame metric. For consistency with notation to be used later, we
have changed $\bar\theta\to\tilde\theta$.}:
\ba
\label{nearh}
ds^2 &=& \frac{2 q }{\sqrt{q_1q_2}} d\tilde{v} d \tilde {r} +
\frac{4L}{\sqrt{q_1q_2}} \tilde{r} d\tilde{v}
d\psi' + \frac{qL^2 }{\sqrt{q_1q_2}} d{\psi'}^2 +
\frac{q_3\sqrt{q_1q_2}}{4} \left( d\tilde{\theta}^2 + \sin^2
\tilde{\theta} d\chi^2 \right) \\
 &&+ \frac{\sqrt{q_1q_2}}{q_3}\left[dz' - \frac{q_3}{2} (1+\cos \tilde{\theta}) d\chi -
\frac{q_1Q_1+q_2Q_2-q_3Q_3+q^3}{2q_1q_2} d\psi' \right]^2 +
\sqrt{\frac{q_1}{q_2}}\; d{\bf z}_4^2.  
\nonumber \ea The near horizon metric \eqn{nearh} is locally the same as that of (oxidized) BMPV. Note however
that the roles of some coordinates, like $\psi'$ and $z'$, are exchanged relative to BMPV. We will revisit this
issue in section \ref{subsec:decoupling}.

The area of the horizon is interpreted as usual as associated to
an entropy. Expressed in terms of the brane numbers, the entropy of the
D1-D5-P black supertube is
\be
S = \frac{\mathcal{A}_H}{4G_5}
=2 \pi
\,\sqrt{N_\mathrm{D1}N_\mathrm{D5}N_\mathrm{P}-
\mathcal{N}_\mathrm{D1}\mathcal{N}_\mathrm{D5}\mathcal{N}_\mathrm{P}
-J_\phi^{2}
-n_\mathrm{D1}n_\mathrm{D5}n_\mathrm{KK}(J_\psi-J_\phi)}\,,
\label{sring}
\ee
where we have defined, in analogy to \reef{eqn:adef},
\beq
\mathcal{N}_\mathrm{D1}=N_\mathrm{D1}-n_\mathrm{D1}n_\mathrm{KK}\,,\quad
\mathcal{N}_\mathrm{D5}=N_\mathrm{D5}-n_\mathrm{D5}n_\mathrm{KK}\,,\quad
\mathcal{N}_\mathrm{P}=N_\mathrm{P}-n_\mathrm{D1}n_\mathrm{D5}\,.
\label{ndef}\eeq
An alternative form is
\beq
S =2 \pi
\,\sqrt{\left[
N_\mathrm{D1}N_\mathrm{D5}-\mathcal{N}_\mathrm{D1}\mathcal{N}_\mathrm{D5}
\right]N_\mathrm{P}+
\left[\mathcal{N}_\mathrm{D1}\mathcal{N}_\mathrm{D5}-
n_\mathrm{KK}(J_\psi-J_\phi)\right]n_\mathrm{D1}n_\mathrm{D5}
-J_\phi^{2}}\,,
\label{sss}\eeq
which suggests the interpretation that the system decomposes into two
sectors, with central
charges $c'=6\left[
N_\mathrm{D1}N_\mathrm{D5}-
\mathcal{N}_\mathrm{D1}\mathcal{N}_\mathrm{D5}\right]$ and
$c''=6\mathcal{N}_\mathrm{D1}\mathcal{N}_\mathrm{D5}$.

It is also worth noting that, in terms of integer brane
numbers equation \reef{quantm} is
\beq
n_\mathrm{D1}\mathcal{N}_\mathrm{D5}+n_\mathrm{D5}\mathcal{N}_\mathrm{D1}-
n_\mathrm{KK}\mathcal{N}_\mathrm{P}=
2 m\; n_\mathrm{D1}n_\mathrm{D5}\,.
\label{quantmn}
\eeq
Note that all dependences on the moduli $g_s$, $R_z/\ell_s$ and
$\ell/\ell_s$ drop out from this equation. It would be interesting to
understand its microscopic origin.

\setcounter{equation}{0}
\section{Non-uniqueness}
\label{sec:nonu}

A supersymmetric black ring solution is completely specified by the
seven dimensionful parameters $Q_i$, $q_i$, and $R$. Such a solution
carries only five independent conserved charges: three gauge charges
proportional to the $Q_i$ and two angular momenta $J_\psi$ and
$J_\phi$. The mass of the solution is not an independent charge
since it is determined by the saturated BPS bound in terms of the
gauge charges. The three dipole charges of the solution,
proportional to the $q_i$, are not conserved charges. Equations
\eqn{MQJ} can be used to eliminate $R$ and one combination of the
dipoles in favour of $J_\psi$ and $J_\phi$. The remaining two
dipoles can still be varied continuously while keeping the conserved
charges fixed. Supersymmetric black rings are therefore not uniquely
determined by the latter, but exhibit infinite non-uniqueness in the
classical theory. In this section we will examine several aspects of
this non-uniqueness.

To simplify the analysis, let us take all gauge charges to be equal,
$Q \equiv Q_i$. The BPS bound then fixes the ADM mass
to be $M=3 \pi Q/(4G_5)$. Since we wish to compare properties of
different rings with the same mass and gauge charges, we define the
dimensionless angular momenta, horizon area and dipole charges as
\bea
 j_{\phi,\psi} =
 \sqrt{\frac{27 \pi}{32 G_5}} \frac{J_{\phi,\psi}}{M^{3/2}} \, ,
 ~~~~~~~
 a_H = \frac{{\cal A}_H}{(G_5 M)^{3/2}}
 \label{jah}
\eea
and
\bea
 \eta_i \equiv \frac{\sqrt{3 \pi}}{2} \frac{q_i}{(G_5 M)^{1/2}}
 = \frac{q_i}{\sqrt{Q}} \, .
\eea
Note from \reef{MQJ} that $j_\psi>j_\phi$. By \reef{Qqq}, the
$\eta_i$'s must satisfy $\eta_i \eta_j \le 1$ for $i \ne j$.
As discussed above, we can eliminate one of the parameters
$\eta_i$. Solving for $\eta_3$ gives
\bea
  \eta_3 = \frac{4\sqrt{2}\, j_\phi - \eta_2 - \eta_1}{1- \eta_2 \eta_1} \, .
\eea
For a regular horizon we need $\eta_i>0$ and hence
$4\sqrt{2}\,j_\phi > \eta_1 + \eta_2$.

It is illustrative to specialize to the case $\eta_1=\eta_2 \equiv
\eta$, \ie equal pitches for the D1 and D5 helices. Substituting in
$\eta_3$ as given above, we find
\bea
a_H &=& \frac{16 \pi^{1/2}}{3^{3/2}}
  \left[
    \frac{(2\sqrt{2}\,j_\phi-\eta)}{1-\eta^2}
     \left(
        \eta \, (3+\eta^2)- 2\sqrt{2}\,j_\phi
       (1+\eta^2)-4\sqrt{2}\,\eta^2 j_\psi
     \right)
  \right]^{1/2}
  \label{aHnonUniq}
\eea
Since $4\sqrt{2}\,j_\phi \ge \eta_1+\eta_2 = 2\eta$ and $\eta \le 1$, we
require
\bea
j_\psi <
j_\psi^\rom{max} \equiv
\frac{1}{4 \sqrt{2} \, \eta^2}\big[ \eta \, (3+ \eta^2)
  - 2 \sqrt{2} \,j_\phi (1+\eta^2)\big] \, .
\label{jpsimax}
\eea
This is just the condition \reef{eqn:noctcs} needed to avoid naked CCCs.
For the BMPV black hole, it is well-known that requiring the spacetime
to be free of naked CCCs imposes an upper bound on the angular
momenta. However, in
the case of BMPV the angular momenta must be equal in magnitude and
there is no equivalent of the non-uniqueness parameter (dipole moment)
$\eta$, so the bound comes out much simpler than \reef{jpsimax}.

\begin{figure}[t]
\begin{center}
\includegraphics[width=8cm]{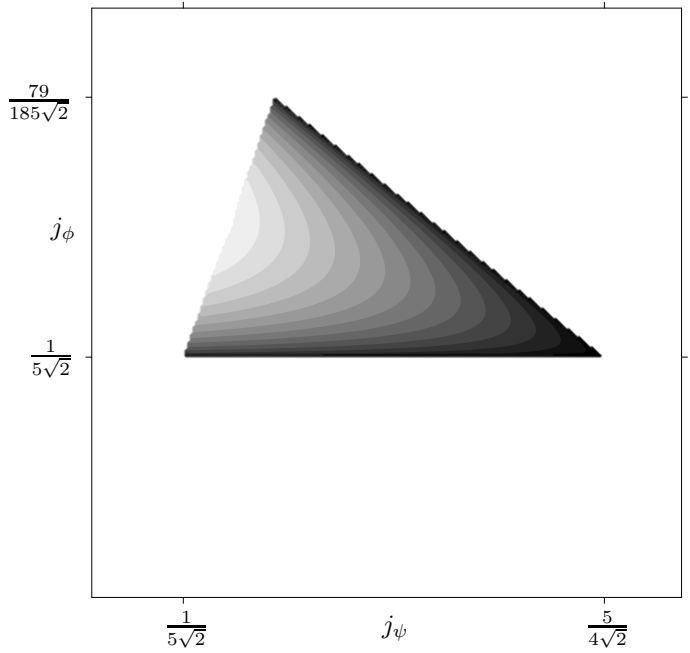}
\begin{picture}(0,0)(0,0)%
\footnotesize{
\put(-86,49){$j_\phi$}%
\put(-42,-4){$j_\psi$}%
\put(-71,-4){$\frac{1}{5\sqrt{2}}$}%
\put(-89,32){$\frac{1}{5\sqrt{2}}$}%
\put(-15,-4){$\frac{5}{4\sqrt{2}}$}%
\put(-92,66){$\frac{79}{185\sqrt{2}}$}%
}
\end{picture}%
\end{center}
\vspace{3mm}
\caption{\small The dimensionless area $a_H$ as a function of $j_\psi$ and
$j_\phi$ for fixed $\eta=0.4$. Note that the third dipole charge is not
held constant, but is determined by the other parameters. Darker
regions correspond to smaller area.
On the left, the triangular region is bounded by the line
$j_\phi=j_\psi$. The lower bound is a consequence of $j_\phi$ being
bounded from below for non-vanishing dipole moment:
$2 \sqrt{2}\,j_\phi \ge \eta$.
The region is bounded on the right by the line determined by
$j_\psi^\rom{max}$ in \reef{jpsimax}. This is a consequence of
requiring that there are no naked CCCs.
The area $a_H$
vanishes at the bottom and right boundaries of the triangular
region. For fixed $j_\phi$, $a_H$ is maximized when $j_\psi \to
j_\phi$.}
\label{fig:aH_fixed_eta}
\end{figure}

In \cite{EEMR}, we studied the supersymmetric black rings obtained
from the general solutions of Section \ref{sec:gensol} by taking all
charges $Q_i$ equal and all dipole moments $q_i$ equal. That
specialized system does not exhibit
non-uniqueness, since the three-parameter solution is specified
uniquely by the conserved charges (the net charge and the two angular
momenta). In particular, we plotted in \cite{EEMR} the horizon area
$a_H$ as a function of $j_\phi$ and $j_\psi$. For the more general case
at hand, we can make such a plot for each value of $\eta$. As an
example, figure \ref{fig:aH_fixed_eta} shows $a_H$ vs $j_\psi$ and
$j_\phi$ for fixed $\eta=0.4$. Note that for non-zero dipole
moments, there are upper and lower bounds on both angular momenta (a
feature present also in non-supersymmetric rings \cite{RE}). It
would be interesting to understand the precise microscopic origin of
these bounds and how they depend on the dipole moments.

The expression \reef{aHnonUniq} for the horizon area $a_H$ illustrates
the non-uniqueness: the net charges are fixed and even when both the
angular momenta are specified, we can still vary $\eta$. In particular,
we can fix $j_\psi$ and $j_\phi$, and plot the entropy $a_H$ as a
function of $\eta$. More generally, we can include both $\eta_1$ and
$\eta_2$. Then we can use one parameter to fix the horizon area $a_H$
and still have another parameter to vary. We conclude that for given net
charges $Q_i$ and angular momenta $j_{\phi,\psi}$, there are infinitely
many supersymmetric black rings with the same horizon area, except for
the black ring that maximizes the area, which is unique. This might
suggest to recover a notion of uniqueness, at least among
supersymmetric black rings, by adding the condition that the solution
have maximum entropy for given conserved asymptotic charges. We
emphasize, however, that this additional requirement is absent from the
traditional notion of black hole uniqueness, and is also known to be
insufficient to distinguish between non-supersymmetric black rings and
black holes of spherical topology \cite{ER}.

\begin{figure}[t]
\centering{
~~~\epsfxsize=16cm\epsfbox{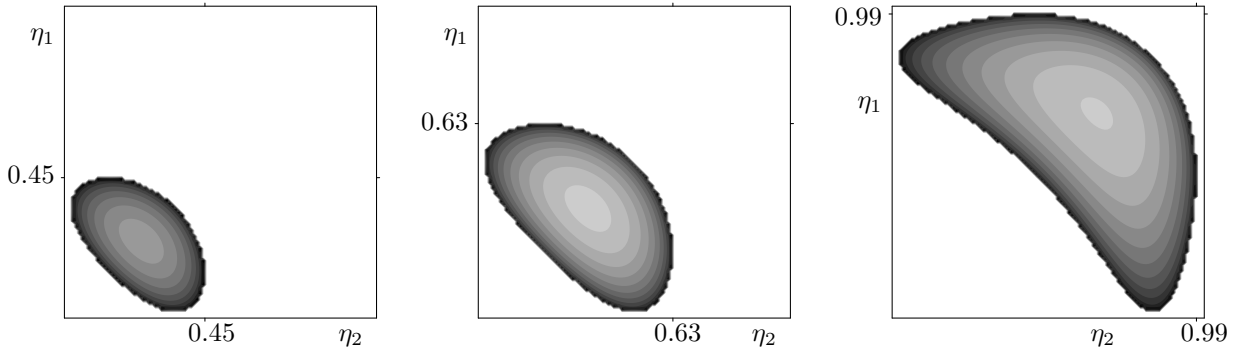}~~~~
}
\begin{picture}(0,0)(0,0)%
\footnotesize{
\put(-146,-1){$0.45$}%
\put(-126,-1){$\eta_2$}%
\put(-167,39){$\eta_1$}%
\put(-170,20){$0.45$}%
\put(-84,-1){$0.63$}%
\put(-71,-1){$\eta_2$}%
\put(-112,39){$\eta_1$}%
\put(-115,27){$0.63$}%
\put(-14,-1){$0.99$}%
\put(-26,-1){$\eta_2$}%
\put(-57,30){$\eta_1$}%
\put(-60,41){$0.99$}%
}
\end{picture}%
\caption{\small Contour plots of the dimensionless area $a_H$
versus $\eta_1$ and $\eta_2$, for fixed values of $j_\phi$ and
$j_\psi$. For all three plots $4\sqrt{2}\,j_\psi = 2.1$, but $j_\phi$
varies for each case: from left to right $4\sqrt{2}\,j_\phi = 0.7$,
$1.$, $1.7$. The plots are symmetric in $\eta_1$ and $\eta_2$, and the darker
regions correspond to smaller area. The regions for which the black
rings exist are bounded by the condition that there be no naked
CCCs. At this boundary, the area $a_H$ vanishes.}
\label{fig:nonuni_aH}
\end{figure}

Figure \ref{fig:nonuni_aH} illustrates the non-uniqueness of
supersymmetric black rings. It shows for fixed values of $j_\phi$ and
$j_\psi$ the horizon area $a_H$ as a function of the dipole
parameters $\eta_1$ and $\eta_2$. The bounds of the covered region
are set by the requirement that there be no naked CCCs.

\vspace{2mm}
Now consider what happens when we uplift the supersymmetric black ring
to a $D=10$ D1-D5-P supertube, as described in section
\ref{sec:d1d5p}. In the quantum theory, the net charges and dipoles
are quantized in terms of the number of branes in the D1-D5-P
configuration. Using \reef{quantN15} and \reef{quantn15}, we find that
the restrictions \reef{Qqq} on the net charges $Q_i$ and dipole
moments $q_i$ become
\be
  N_\rom{D5} \geq n_\rom{D5}  \, n_\rom{KK} \,,\quad
  N_\rom{D1} \geq n_\rom{D1}  \,  n_\rom{KK} \,,\quad
  N_\rom{P}  \geq n_\rom{D1}  \,  n_\rom{D5} \,.
\ee
So the number of D1- and D5-branes and units of momenta restrict the
number of dipole branes and KK monopoles. This shows that upon
quantization of the charges, the non-uniqueness becomes finite (but still very large).

\setcounter{equation}{0}
\section{Particular cases and limits}
\label{sec:cases}

In this section we study various limits of the supersymmetric black
ring solution. First in subsection \ref{bmpv}, we consider the limit
$R\to 0$ where the solution reduces to the BMPV black hole. In the
infinite radius limit $R\to \infty$, the black ring becomes a black
string in five dimensions. We show in subsection \ref{infR} that in
this limit our solution reproduces the black string metric found in
\cite{bena}. Subsection \ref{simpler} contains special cases
of the general black ring solution: one is the original
two-charge supertube solution \cite{MT,EMT}, the other the three-charge
solution with only two nonzero dipole moments.
Finally, in subsection \ref{subsec:decoupling}, we study the
decoupling limit relevant for the AdS/CFT correspondence.

\subsection{BMPV black hole}
\label{bmpv}

\begin{figure}[th]
\begin{picture}(0,0)(0,0)
\footnotesize{
\put(46,-3){$2/27$}
\put(77,-3){$1/8$}
\put(60,-6){$j^2$}
\put(-12,33){$a_H$}
\put(-8,48){$\frac{16}{3}\sqrt{\frac{\pi}{3}}$}
\put(-8,16){$\frac{16}{9}\sqrt{\frac{\pi}{3}}$}
}
\end{picture}
\centering{\epsfxsize=8cm\epsfbox{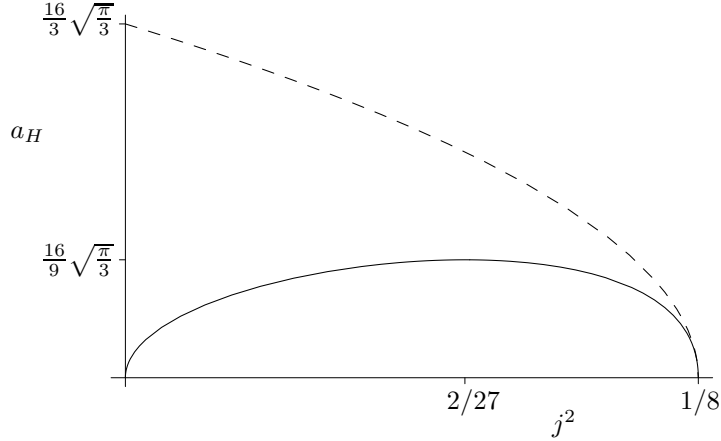}}
\vskip.8cm
\caption{\small Area of the BMPV black hole (dashed) and limit $R\to 0$
of the area of the supersymmetric black ring (solid), vs.\ spin$^2$, for
fixed mass (in terms of the variables $a_H$ and $j$ in \reef{jah}). For
simplicity the three charges and dipoles are set equal, $Q_i=Q$, $q_i=q$
so these are solutions of minimal $D=5$ supergravity.}
\label{fig:comparison}
\end{figure}

Consider the solution in the $(\rho,\Theta)$ coordinates of
\reef{rhoTheta}--\reef{rhoThetabase}\footnote{The limit can equally well
be taken in the $(r,\theta)$ coordinates \reef{rtheta}: when $R=0$ one
has $\rho=r$,
$\Theta=\theta$.}.
If we set $R= 0$ we find $H_i = 1+ Q_i/\rho^2$ and
\be
  \omega_\phi = - \frac{4G_5J}{\pi} \frac{\cos^2{\Theta}}{\rho^2}
  \, ,~~~~
  \omega_\psi = - \frac{4G_5J}{\pi} \frac{\sin^2{\Theta}}{\rho^2} \, ,~~~~ A^i = H_i^{-1} (dt +\omega) \, ,
\ee
where
\be
  J= \frac{\pi}{8G_5}
  \left[ q_1 Q_1 + q_2 Q_2 + q_3 Q_3 - q_1 q_2 q_3 \right] \, .
\ee
This is the BMPV black hole with three independent charges $Q_i$ and
angular momenta $J_\psi = J_\phi = J$ \cite{bmpv}. Note that the parameters $q_i$ have become
redundant since they enter the solution only through the angular momentum $J$. In particular, they no longer appear in $A^i$ and therefore no longer have the interpretation of dipole charges.
The horizon is located at $\rho \to 0$. Topologically the horizon is a
three-sphere. The horizon area is
\be
  \mathcal{A}_\rom{BMPV} =
  2 \pi^2 \sqrt{Q_1 Q_2 Q_3 - \left(\frac{4G_5J}{\pi}\right)^2} \, .
\ee
This is \emph{not} the $R\to 0$ limit of the horizon area of the
supersymmetric black ring \reef{ringarea},
\beq
\lim_{R\to 0}  \mathcal{A}_\rom{ring} =
  2 \pi^2 \sqrt{Q_1 Q_2 Q_3 -\mathcal{Q}_1\mathcal{Q}_2\mathcal{Q}_3-
\left(\frac{4G_5J}{\pi}\right)^2} \,,
\eeq
which is always smaller than that of the BMPV black hole with the same
asymptotic charges, except possibly when both areas vanish. The areas are
compared in fig.~\ref{fig:comparison} for the particular case of equal
charges $Q_i=Q$ and dipoles $q_i=q$, \ie for the solutions of minimal
supergravity.

A clue to a (macroscopic) understanding of this effect follows by
considering an analogy to a two-center extremal Reissner-Nordstrom (RN)
solution. Observe that if we take a diameter section of the black ring
solution, at constant $\psi$ and $\psi +\pi$, then the resulting
four-dimensional geometry contains two infinite throats and therefore is
similar to the geometry of two extremal black holes. The limit in which
the inner radius of the ring shrinks to zero is in this sense analogous
to the process in which the two centers of the extremal black holes are
taken to coincide. If two RN extremal black holes, each of charge $Q$,
are separate, the total mass is $2Q$ and the total area is $2(4\pi
Q^2)$. When their centers coincide the mass is still $2Q$ but the
entropy jumps to $4\pi (2Q)^2$. So the limit of zero separation is
discontinuous, and indeed the topology of the solution changes. This
same effect occurs for the black ring: the area of the solution with
$R=0$, \ie the BMPV black hole, is larger than the limit $R \to 0$ of
the ring area, and the topology of the two solutions are
different.

\subsection{Infinite radius limit}
\label{infR}

In this limit we take $R\to\infty$, but keep the charges per
unit length along the tube ($2\pi R$) finite by defining finite linear
densities (the factors $\Omega_n$ account for different
dimensionalities of spheres for integration)
\beq
{\bar Q_i}=\frac{Q_i}{2\pi R}\frac{2\Omega_3}{\Omega_2}=\frac{Q_i}{2R}
\eeq
We keep $q_i$ fixed and define finite coordinates $\bar r$, $\bar \theta$,
$\eta$ by\footnote{These $\bar r$, $\bar \theta$ are the same as
introduced in the near-horizon study of appendix~\ref{app:through}.}
\beq
\bar r=-R/y\,,\quad \cos\bar \theta=x\,,\quad \eta=R\psi \, .
\eeq
Then we get
\beqa
  H_1 &\to&  1+ \frac{\bar{Q}_1}{\bar r} + \frac{ q_2 q_3}{4 \bar r^2}\,, \\[2mm]
  \omega_\psi d\psi &\to& -\left(
     \frac{q_1+q_2+q_3}{2 \bar r}
     +\frac{q_1 \bar{Q}_1+q_2 \bar{Q}_2+q_3 \bar{Q}_3}{4 \bar r^2}
     +\frac{ q_1q_2q_3}{8 \bar r^3}\right) d\eta\,, \\[2mm]
  \omega_\phi &\to& 0 \, ,
\eeqa
with $H_2$ and $H_3$ given by permutations of (123). With this
we reproduce the metric for the `flat supertube' in \cite{bena}.

\subsection{Simpler supertubes}
\label{simpler}

\subsubsection{Two charges and one dipole}

The original supergravity supertube solutions in $D=6$
\cite{globalads3,LM1,EMT} correspond to setting $Q_3=0$ and therefore
$q_1=q_2=0$. In this case $J_\phi=0$ and $J_\psi=(\pi/4G_5) R^2 q_3$.
The bound \reef{eqn:noctcs} is trivially saturated, but this
inequality is only sufficient to eliminate CCCs when $q_i>0$. In the
present case, it is easy to use the results of Appendix \ref{app:noctcs}
to see that the necessary and sufficient condition for the absence of
CCCs is\footnote{In this case, the first two lines of equation
\reef{eqn:delta} vanish, as does the second term on the third line, so
the leading order (cubic) behaviour as $y \rightarrow -\infty$ comes
from $Y$. Demanding $Y>0$ gives this inequality.} $q_3^2 \le Q_1 Q_2 /
R^2$, which agrees precisely with the worldvolume analysis of these
supertubes in \cite{MT,EMT}.

\subsubsection{Three charges and two dipoles}

The solution with $q_3=0$ and $Q_{1,2,3}\neq 0$, $q_{1,2}\neq 0$, is more
complicated than the previous one in that the BPS equations that it
solves are non-linear and therefore the functions $H_i$ are not
harmonic. There are also two independent angular momenta.

The solution can be interpreted as a helical D1-D5 string carrying
momentum in the direction of the axis of the helix, along which it is
smeared, but this time the KK monopole tube is absent. It is different
than the D1-D5-P gyrating strings of \cite{homa}, since it has two
independent angular momenta and the area is always zero. In fact, now
there is a naked curvature singularity at $y=-\infty$ where
$R_{\mu\nu}R^{\mu\nu}$ diverges.

Absence of causal anomalies imposes again constraints on the parameters.
Equation \reef{eqn:noctcs} reduces to $-(q_1 Q_1 - q_2 Q_2)^2 \ge 0$ so
we require
\be
q_1 Q_1 =q_2 Q_2\,.
\label{qQ}\ee
In terms of quantized brane numbers
\reef{quantN15}, \reef{quantn15}, this equation becomes
\beq
\frac{N_\mathrm{D1}}{n_\mathrm{D1}}=\frac{N_\mathrm{D5}}{n_\mathrm{D5}}\,,
\label{32noctc1}
\eeq
which means that the D1 and D5 helices have the same pitch and can
therefore bind to form a D1-D5 helix.

From Appendix C, it is now easy to see that the necessary and
sufficient condition for absence of CCCs is that the radius $R$ be
bounded above like\footnote{
The first line of equation \reef{eqn:delta} vanishes, as does the
second term on the second line, so the leading (quartic) behaviour as
$y \rightarrow -\infty$ comes from $X$. Demanding $X>0$ gives this
inequality. It is then easy to see that $Y>0$ so the remaining terms
in \reef{eqn:delta} are positive.}
\be
R^2\leq \frac{(Q_3-q_1q_2)Q_1Q_2}{q_1q_2(Q_1 + Q_2)}
\,.
\label{QqR}\ee

Eq.~\reef{32noctc1} is the precise T-dual (in the $z$ direction) to a
constraint found for supertubes with D0-D4-F1 charges and D2-D6 dipoles
using the worldvolume theory of D6-branes
\cite{benakraus}\footnote{Eq.~\reef{qQ} was also recovered in the
infinite radius limit in \cite{bena}.}. Eq.~\reef{QqR} is very similar
to, but not exactly the same as, another equation derived in
\cite{benakraus} for the same system. We will return to this point in
Sec.~\ref{sec:sugravsworld}.

It is perhaps surprising that despite having the three D1-D5-P
charges, the area of this solution always vanishes. The apparent reason
is that near the core at $y\to-\infty$ the
solution is mostly dominated by terms involving the dipole moments $q_i$.
There is a curvature singularity at the core and the moduli also blow
up there.
However, it will be shown in \cite{EEF} that this solution admits
thermal deformations, \ie there exists a family of non-extremal
solutions with regular horizons, which in the extremal limit reduce to
the supersymmetric solution with three charges and two dipoles.

\bigskip

One can easily check that the solutions with three charges and one
dipole, as well as the ones with two charges and two dipoles, always
possess CTCs near $y\to-\infty$. This parallels the fact that these
supertubes do not have a sensible Born-Infeld description
\cite{benakraus,bena}.

\subsection{Decoupling limit}
\label{subsec:decoupling}

In the decoupling limit of the D1-D5-P solution we send
$\alpha'=\ell_s^2\to 0$ and keep the string coupling $g_s$ fixed, in
such a way that the geometry near the core decouples from the
asymptotically flat region. Our solution contains several more
parameters than previous D1-D5-P configurations, so we shall describe
this limit in some detail.

We work with the coordinates
$(r,\theta)$ defined in \reef{rtheta}. Since we want to keep fixed the
energies (in string units) of the excitations that live near the core,
then $r/\alpha'$ and $R/\alpha'$ must remain finite. We are also interested in
a regime where the size of the $z$ direction, $R_z$, is large compared to
$\ell_s$, so that winding modes can be ignored and the momentum
modes are the lowest excitations. So when we take $\alpha'\to 0$, $R_z$
will be fixed, and then also $q_3$ as a result of
\reef{q3KK}. Further, we take the $T^4$ length scale
$\ell \sim \ell_s$ so that the energy scale of both momentum and winding
modes on the $T^4$ is large. Finally, we keep
the string coupling fixed and also keep the number of branes and units
of momentum fixed. Using \reef{quantN15} and \reef{quantn15} the
limiting solution is obtained taking $\alpha'\to 0$ while
\beqa
r/\alpha'\,,&&\qquad Q_{1,2}/\alpha'\,,\qquad Q_3/{\alpha'}^2\,,\nn
R/\alpha'\,,&&\qquad q_{1,2}/\alpha'\,,\qquad q_3\,
\label{declimit}\eeqa
are held fixed.
Then the length scales in the supergravity solution are arranged as
\beq
r\sim R\sim \sqrt{Q_3}\sim \sqrt{q_1 q_2}\ll (Q_1Q_2)^{1/4}\sim
\sqrt{q_3}(q_1q_2)^{1/4}
\label{decrange}\eeq
with $Q_1\sim Q_2$ and $q_1\sim q_2$.
Recall from \reef{supertube} that $Q_1$, $Q_2$, and $Q_3$ label D5, D1,
and momentum charge, respectively, and $q_1$, $q_2$, and $q_3$ correspond
to the dipole charges of respectively d1- and d5-branes and the
Kaluza-Klein monopoles making up the supertube. Observe also that
$\mathcal{Q}_i$ scales like $Q_i$.

After rescaling the metric and the gauge fields $A^1$ and $A^2$ by an
overall factor of $\alpha'$, we obtain a new solution of IIB
supergravity. This decoupled solution has the same form as
\reef{eqn:2bmetric}, \reef{2bfields}, \eqn{eqn:5dsol} with, in the
coordinates of \reef{rthetabase},
\ba
H_{1,2}&=&\frac{Q_{1,2}}{\Sigma}-\frac{q_{2,1}q_3R^2\cos 2\theta}{\Sigma^2}\,,\nn
H_3&=&1+\frac{Q_3}{\Sigma}-\frac{q_1q_2R^2\cos 2\theta}{\Sigma^2}\,,
\ea
\beqa
  \omega_\psi &=&
    -q_3\frac{R^2\sin^2\theta}{\Sigma}
    - \frac{(r^2+R^2) \sin^2{\theta}}{\Sigma^2}
    \left(\frac{4G_5 J_\phi}{\pi}- q^3\frac{R^2\cos 2\theta}{\Sigma}\right)
  \, .\nonumber
\eeqa
while $\omega_\phi$ and $A^i$ remain as in \reef{omegarth} and
\reef{Arth}, and $J_\phi$ is as in \reef{MQJ}. We can gauge-transform
$A^3 \to A^3 -dt$, \ie $z\to z-t$, so that $A^3_t=H_3^{-1}-1$ vanishes
at $r\to \infty$. It is apparent that this decoupling limit amounts to
the familiar procedure of ``removing the 1's" from the functions
$H_{1,2}$ associated to the D5 and D1 branes, with the first term of
$\omega_\psi$ modified so the result remains a solution of the field
equations.

The decoupling limit is not in general the same as the near-horizon
limit \reef{nearh} analyzed earlier. In the near-horizon limit $r$ is
taken to be much smaller than $R$ and indeed than any other scale in the
system, so one covers only a small region of the decoupled
solution. Also, the near-horizon limit in \reef{nearh} exists for any
black ring, whereas here we are restricting the parameters to the ranges
in \reef{decrange}. These differences between the two limits are in fact
also present for BMPV \cite{cvla}.
The two limits nevertheless commute, so the new solution has a regular
horizon of finite area. After an appropriate rescaling, $L$ is the same as
\reef{bigL}, when expressed in terms of physical quantities only. A
slight difference is hidden, though,
in the fact that in the decoupling limit $J_\psi-J_\phi$ changes to
\beq
(J_\psi-J_\phi)_{\rm{decoupled}}=\frac{\pi}{4G_5}R^2 q_3\,.
\eeq
This is simply a consequence of the restrictions \reef{declimit} on the
relative values of the parameters, which imply in particular that $q_3\gg
q_{1,2}$. As a consequence, rings with $q_3\sim q_{1,2}$ are not
expected to be fully captured by the dual CFT description of D1-D5
systems.

At asymptotic infinity, $r\to\infty$, the metric becomes (omitting the $T^4$ factor)
\beq
ds^2\to \frac{r^2}{\sqrt{Q_1Q_2}}(-dt^2+dz^2)+\sqrt{Q_1 Q_2}\;\frac{dr^2}{r^2}
+\sqrt{Q_1 Q_2}(d\theta^2+\sin^2\theta d\psi^2+\cos^2\theta d\phi^2)\,,
\label{asympads}\eeq
so we recover the asymptotic geometry of global (as $z$ is periodic)
AdS$_3$ times $S^3$, both with equal radius
\beq
\ell_\infty=(Q_1Q_2)^{1/4}\,.
\eeq
For certain particular values of the parameters we get solutions which
are {\it everywhere} locally AdS$_3\times S^3$. This happens when $R=0$,
\ie BMPV in the decoupling limit \cite{cvla}, and also for two-charge
D1-D5 supertubes that saturate the CTC bound $q_3^2 R^2=Q_1 Q_2$. The
decoupling limit of the latter is global AdS$_3$ times a rotating $S^3$,
with a conical singularity if $n_\mathrm{KK}>1$
\cite{globalads3}.

However, in general our solution is (locally) the product space AdS$_3\times
S^3$ only at asymptotic infinity and near the horizon. In the latter
region this occurs in a rather unusual manner. In order to find the
near-horizon limit of the decoupling solution in $(r,\theta)$
coordinates, near $r= 0$, $\theta= \pi/2$, we take the gauge in which
$A^3_t=H_3^{-1}$ so we are in a frame
that corotates with the horizon, and define
\beq
r^2=\epsilon \tilde r L \cos^2\frac{\tilde\theta}{2}\,,\qquad
R^2\cos^2\theta=\epsilon \tilde r L
\sin^2\frac{\tilde\theta}{2}\,,\qquad t=\tilde t/\epsilon\,.
\eeq
Sending $\epsilon\to 0$, the geometry
that results is the same as \reef{nearh}, but now in coordinates
that cover only the region outside the horizon,
\ba
 ds^2 &=& \frac{2L}{\sqrt{q_1q_2}}\, \tilde{r}\, d\tilde{t}\,
d\psi + \frac{qL^2 }{\sqrt{q_1q_2}}\, d{\psi}^2
+\frac{q_3\sqrt{q_1q_2}}{4} \frac{d\tilde{r}^2}{\tilde{r}^2}\\
&&+ \frac{\sqrt{q_1q_2}}{q_3}\left(dz - \frac{q_3}{2}
(1+\cos\tilde{\theta}) d\chi \right)^2+ \frac{q_3\sqrt{q_1q_2}}{4}
\left( d{\tilde{\theta}}^2 + \sin^2\tilde{\theta} d\chi^2 \right)
\,,  \nonumber
\label{nearh2}\ea
where we use $\chi=\phi-\psi$ and, for simplicity, we assume
\reef{quantm} is satisfied so that we can use a shift $z\to z +
m\,R_z\psi$ to bring the geometry to a product of two factors. The first
factor is locally isometric to AdS$_3$ with radius
\beq
\ell_\mathrm{nh}=\sqrt{q_3}(q_1q_2)^{1/4}\,.
\eeq
This factor is  globally the same as the near-horizon limit of an
extremal BTZ black hole with mass and spin
\beq
M_\mathrm{BTZ}=2L^2/q^2\,,\qquad J_\mathrm{BTZ}=
M_\mathrm{BTZ}\ell_\mathrm{nh}\,,
\eeq
and horizon at $\tilde r=0$. The second factor is
the quotient space $S^3/Z_{n_\mathrm{KK}}$, with the same radius
$\ell_\mathrm{nh}$.

The appearance of AdS$_3\times S^3$ near the horizon is very different
from the factorization into AdS$_3\times S^3$ in the asymptotic region
\reef{asympads}. The AdS$_3$ near the horizon spans the coordinates
$(\tilde{t},\tilde{r},\psi)$ whereas near the asymptotic boundary it
spans $(t,r,z)$ ---the relation between $t$ and $\tilde{t}$ just amounts
to the redshift near the horizon, but the other coordinates are not
simply related.

The direction in which the near-horizon geometry rotates is $\psi$. In
contrast, in the decoupling limit of BMPV, the near-horizon geometry
rotates in the $z$-direction and arises from the linear momentum in this
direction. Here the rotation of the near-horizon geometry arises, in a
sense, from the rotation of the ring, but there is no simple
relationship between $J_\psi$ and $J_\mathrm{BTZ}$. Furthermore, the
radii of the two AdS$_3$'s are different
\be
\label{eqn:uvir}
  \ell_\infty >  \ell_\mathrm{nh}.
\ee
The curvature of the solution near the horizon is not controlled by the
net D1 and D5 charges but instead by the dipole charges $q_i$. As a
consequence, the simple argument for the statistical calculation of the
entropy of the BTZ black hole, from the charges under the Virasoro algebra of
diffeomorphisms at the boundary of AdS$_3$ \cite{str97}, does not seem
to apply easily to the computation of the entropy of the ring.

So the full decoupling solution interpolates in a highly non-trivial way
between two different factorizations of the six-dimensional solution,
both of which are locally of the form AdS$_3\times S^3$. Between these
two limiting regions, the solution has non-vanishing Weyl curvature and
is generically quite complicated. Indeed, already when the first
subleading terms near $r\to\infty$ are considered, the geometry does not
factorize.

This decoupled solution must admit a dual description in terms of an
ensemble of supersymmetric states of the dual CFT. It should be very
interesting to identify and count the degeneracy of these states to
reproduce the entropy of the black ring. Given the two limiting AdS
geometries, it might be useful to view the solution as dual to a
renormalization group flow, with equation \eqn{eqn:uvir} implying that
the central charge is greater in the UV than in the IR, as expected from
the $c$-theorem.

\setcounter{equation}{0}
\section{Worldvolume Supertubes and K\"ahler Calibrations}
\label{sec:worldv}

Consider three M5-branes intersecting as in the array \reef{intersection}, with the $\psi$-circle replaced by a
curve $C$ in the $\bbe{4}$ space transverse to the M2-branes. If $C$ is a straight line, then the worldspace of
the three M5-branes may be described as that of a single M5-brane with (in general) non-singular worldspace $S
\times C$, where $S$ is a K\"ahler-calibrated surface of degree four (that is, of complex dimension two) embedded
in the $\bbc{3} = \bbe{6}$ space spanned by the $z^1, \ldots, z^6$ coordinates. This configuration preserves 1/8
of the thirty-two supersymmetries of the M-theory Minkowski vacuum. Here we will show that an M5-brane with
worldspace $S \times C$ and appropriate fluxes of the worldvolume three-form $H$ also preserves 1/8 of the
supersymmetries (albeit a different set) for any {\it arbitrary} curve $C$ in $\bbe{4}$. If $C$ is closed then
this configuration carries no net M5-brane charges, but only three M2-brane charges and three M5-brane dipoles,
and thus provides the first worldvolume description of a three-charge supertube in which the three dipoles are
visible. We call this a {\it calibrated supertube}. In fact, we will show that {\it any} K\"ahler calibration (of
appropriate degree to be interpreted in terms of M5-branes) gives rise to a calibrated supertube in a similar
manner. It would be interesting to investigate the relationship between supertubes and other types of calibrations
(SLAG calibrations, exceptional calibrations, etc..), with calibrations in more general backgrounds, and with
calibrations corresponding to non-static branes.

To avoid any confusion, it is worth mentioning from the start a
limitation of the worldvolume description we are about to present.
This is the fact that, although the configurations we will construct
do carry multiple charges and dipoles globally, they may be locally
regarded as standard two-charge/one-dipole supertubes. Globally,
therefore, they may be viewed as resolved junctions of standard
supertubes, in the same sense that certain calibrations can be
viewed as resolved junctions of M5-branes. One very concrete
manifestation of this limitation is that the worldvolume description
does not capture the second angular momentum visible in the
supergravity description.

Despite this limitation, it is remarkable that such non-singular
junctions of supertubes can preserve supersymmetry,\footnote{{\it
Non}-supersymmetric supertube junctions have been previously studied
in \cite{BL}.} and we regard the construction below as a first step
towards a more sophisticated worldvolume description of three-charge
black supertubes.

\subsection{K\"ahler calibrations and intersecting M5-branes}

We begin by reviewing a few facts about K\"ahler calibrations. We
follow closely the discussion in \cite{GP,GLW}. Let
$u^j = z^{2j-1} + i z^{2j}$, with $j=1, \ldots, n$, be complex coordinates
on $\bbc{n} = \bbe{2n}$, with metric
\be
ds^2 = \sum_{j=1}^n du^j \, d\bar{u}^j = \sum_{j=1}^{2n} dz_j^2 \,,
\ee
and
\be
\calh = \frac{i}{2} \sum_{j=1}^n du^j \wedge d\bar{u}^j =
\sum_{j=1}^n dz^{2j-1} \wedge dz^{2j}
\label{calh}
\ee
the associated K\"ahler two-form. Then
\be
\Psi = \frac{1}{p!} \, \calh^p
\label{Psi}
\ee
is a calibration of degree $2p$ in $\bbc{n}$ ($p \leq n$) associated
to the group $SU(n)$ \cite{HL}. This means that the volume of any
(hyper)surface $S \subset \bbe{2n}$ of dimension $2p$ is bounded
from below as
\be
\int_S d^{2p} \xi \, \sqrt{\det g} \geq \int_S \, \Psi \,,
\ee
where $\xi$ are coordinates on $S$, $g$ is the induced metric on
$S$, and a pull-back of $\Psi$ onto $S$ is understood.
If this bound is saturated, the surface $S$ is said to be calibrated
by $\Psi$, or K\"ahler calibrated (since $\Psi$ is constructed from
the K\"ahler form).
All complex surfaces in $\bbc{n}$ are K\"ahler calibrated \cite{GP}.
Since $S$ minimizes its volume within its homology class,
a static M5-brane with worldspace $S \times \bbe{5-2p}$ is a
solution of the M5-brane equations of motion. Moreover, any two
tangent (hyper)planes to $S$ are related by an $SU(n) \subset
SO(2n)$ rotation, from which it follows that a fraction $1/2^n$ of
supersymmetry is preserved.

The K\"ahler calibrations of interest here are those that have an
interpretation in terms of intersecting M5-branes, namely those with
$n=2,3,4$. For $n=2$ there is only an $SU(2)$ calibration of degree two,
corresponding to a Riemann surface $S$ embedded in
$\bbc{2}$. An M5-brane with worldspace $S\times \bbe{3}$ can be
interpreted as the intersection of two M5-branes
\be
\begin{array}{rcccccccl}
\mbox{M5:}\,\,\, & 1 & 2 & \_ & \_ & 5 & 6 & 7 & \, \\
\mbox{M5:}\,\,\, & \_ & \_ & 3 & 4 & 5 & 6 & 7 & \,,
\end{array}
\label{2}
\ee
where the $\bbc{2}$ space corresponds to the 1234-directions.

For $n=3$ there are two relevant calibrations, of degrees two and
four. The first one corresponds to a Riemann surface $S$ in
$\bbc{3}$. An M5-brane with worldspace $S \times \bbe{3}$
preserves 1/8-supersymmetry and can be interpreted as describing the
triple intersection
\be
\begin{array}{rcccccccccl}
\mbox{M5:}\,\,\, & 1 & 2 & \_ & \_ & \_ & \_ & 7 & 8 & 9 &  \, \\
\mbox{M5:}\,\,\, & \_ & \_ & 3 & 4 & \_ & \_ & 7 & 8 & 9 & \, \\
\mbox{M5:}\,\,\, & \_ & \_ & \_ & \_ & 5 & 6 & 7 & 8 & 9 & \,.
\end{array}
\label{3a}
\ee
The $SU(3)$ calibration of degree four corresponds to a surface
$S$ of complex dimension two embedded in $\bbc{3}$. An M5-brane with
worldspace $S \times \bbe{}$ can be interpreted as the triple
intersection
\be
\begin{array}{rcccccccl}
\mbox{M5:}\,\,\, & \_ & \_ & 3 & 4 & 5 & 6 & 7& \, \\
\mbox{M5:}\,\,\, & 1 & 2 & \_ & \_ & 5 & 6 & 7 & \, \\
\mbox{M5:}\,\,\, & 1 & 2 & 3 & 4 & \_ & \_ & 7 & \,.
\end{array}
\label{3b}
\ee
In both $SU(3)$ cases, the $\bbc{3}$ space corresponds to the
123456-directions. As mentioned above, the second case corresponds
to the M5 intersection in \eqn{intersection} with the circle
replaced by a line.

Finally, the only $SU(4)$ calibration that has an interpretation in
terms of M5-branes is that of degree four, which corresponds to a
surface $S$ of complex dimension two embedded in
$\bbc{4}$. \footnote{An
$SU(4)$ calibration of degree two has an interpretation as an
intersection of four M2-branes, and one of degree six as an
intersection of a number of D6-branes.} An M5-brane with worldspace
$S \times \bbe{}$ can be interpreted as the sixtuple intersection
\be
\begin{array}{rcccccccccl}
\mbox{M5:}\,\,\, & 1 & 2 & 3 & 4 & \_ & \_ & \_ & \_ & 9 & \, \\
\mbox{M5:}\,\,\, & 1 & 2 & \_ & \_ & 5 & 6 & \_ & \_ & 9 & \, \\
\mbox{M5:}\,\,\, & 1 & 2 & \_ & \_ & \_ & \_ & 7 & 8 & 9 & \, \\
\mbox{M5:}\,\,\, & \_ & \_ & 3 & 4 & 5 & 6 & \_ & \_ & 9 & \, \\
\mbox{M5:}\,\,\, & \_ & \_ & 3 & 4 & \_ & \_ & 7 & 8 & 9 & \, \\
\mbox{M5:}\,\,\, & \_ & \_ & \_ & \_ & 5 & 6 & 7 & 8 & 9 & \,.
\end{array}
\label{4}
\ee

Although in all arrays above we have displayed the M5-branes as
being orthogonal, this need not be the case; they can intersect at
arbitrary $SU(n)$ angles, which are encoded in $S$. Let us
illustrate this for the $SU(3)$ calibration of degree four,
represented by \eqn{3b}. The surface $S$ can generally be specified
as the locus $F(u_1, u_2, u_3)=0$, where $F$ is a holomorphic
function. The choice of $F$ determines the $SU(3)$ angles between
the three M5-branes, which arise at asymptotic regions of $S$. As an
example, consider $F=f_1 f_2 f_3 - c$, where
$f_i = \sum_{j=1}^3 a_i^j \, u^j$ are linear functions and
$a_i^j$ and $c$ are constants. The induced metric on $S$
has three asymptotically flat regions that can be identified with
the three M5-branes. A simple way to determine these is to set
$c=0$. In this case the locus $F=0$ consists of three complex
planes, $f_1=0$, $f_2=0$ and $f_3=0$, the angles between them being
determined by the constants $a_i^j$. This corresponds to a singular
intersection of three M5-branes that extend along these three
planes. Setting now $c \neq 0$ smooths out the intersection and
hence allows the entire complex two-surface to be interpreted as a
single M5-brane, but does not alter the orientations of the
asymptotic regions, which therefore can still be identified with
three distinct M5-branes.

Despite the fact that the arrays above do not necessarily specify
the $SU(n)$ angles between the intersecting branes, each array is
useful in summarizing the number of participating branes and the set
of supersymmetries preserved by each intersection. For example, the
configuration represented by the array \eqn{2} preserves
1/4-supersymmetry, corresponding to the Killing spinors $\eta$
subject to the constraints
\be
\Gamma_{012567} \, \eta = \eta \sac
\Gamma_{034567} \, \eta = \eta \,,
\ee
each of them being associated to one of the M5-branes. Similarly,
the configuration represented by \eqn{4} preserves
1/16-supersymmetry, corresponding to any four of the six M5-branes
(the two projectors associated to any two of the M5-branes are
implied by the those of the other four).

\subsection{Calibrated supertubes}

We are now in a position to show that each of the K\"ahler
calibrations above gives rise, through turning on appropriate
worldvolume fluxes, to a calibrated supertube. Choosing
$S = \bbe{2} \times S'$, with  $S'$ a Riemann surface,
in the $SU(3)$ and $SU(4)$ calibrations of degree four, represented
by the arrays \eqn{4} and \eqn{3b}, we recover the
$SU(2)$ and $SU(3)$ calibrations of degree two, represented by the
arrays \eqn{2} and \eqn{3a}, respectively. The two degree-two
calibrations can therefore be regarded as `degenerate' cases of the
two degree-four calibrations, and for this reason we will only
discuss the latter two. These give rise to the
three-charge/three-dipole $SU(3)$ calibrated supertube
\be
\begin{array}{rcccccccl}
Q_1 \,\, \mbox{M2:}\,\,\, & 1 & 2 & \_ & \_ & \_ & \_ & \_ & \, \\
Q_2 \,\, \mbox{M2:}\,\,\, & \_ & \_ & 3 & 4 & \_ & \_ & \_ & \, \\
Q_3 \,\, \mbox{M2:}\,\,\, & \_ & \_ & \_ & \_ & 5 & 6 & \_ & \, \\
q_1 \,\, \mbox{m5:}\,\,\, & \_ & \_ & 3 & 4 & 5 & 6 & \sigma & \, \\
q_2 \,\, \mbox{m5:}\,\,\, & 1 & 2 & \_ & \_ & 5 & 6 & \sigma & \, \\
q_3 \,\, \mbox{m5:}\,\,\, & 1 & 2 & 3 & 4 & \_ & \_ & \sigma & \, \\
\end{array}
\label{super3}
\ee
and to the four-charge/six-dipole $SU(4)$ calibrated supertube
\be
\begin{array}{rcccccccccl}
Q_1 \,\, \mbox{M2:}\,\,\, & 1 & 2 & \_ & \_ & \_ & \_ & \_ & \_ & \_ & \, \\
Q_2 \,\, \mbox{M2:}\,\,\, & \_ & \_ & 3 & 4 & \_ & \_ & \_ & \_ & \_ & \, \\
Q_3 \,\, \mbox{M2:}\,\,\, & \_ & \_ & \_ & \_ & 5 & 6 & \_ & \_ & \_ & \, \\
Q_4 \,\, \mbox{M2:}\,\,\, & \_ & \_ & \_ & \_ & \_ & \_ & 7 & 8 & \_ & \, \\
q_1 \,\, \mbox{m5:}\,\,\, & 1 & 2 & 3 & 4 & \_ & \_ & \_ & \_ & \sigma & \, \\
q_2 \,\, \mbox{m5:}\,\,\, & 1 & 2 & \_ & \_ & 5 & 6 & \_ & \_ & \sigma & \, \\
q_3 \,\, \mbox{m5:}\,\,\, & 1 & 2 & \_ & \_ & \_ & \_ & 7 & 8 & \sigma & \, \\
q_4 \,\, \mbox{m5:}\,\,\, & \_ & \_ & 3 & 4 & 5 & 6 & \_ & \_ & \sigma & \, \\
q_5 \,\, \mbox{m5:}\,\,\, & \_ & \_ & 3 & 4 & \_ & \_ & 7 & 8 & \sigma & \, \\
q_6 \,\, \mbox{m5:}\,\,\, & \_ & \_ & \_ & \_ & 5 & 6 & 7 & 8 &
\sigma & \,.
\end{array}
\label{super4}
\ee
In the two arrays above, we have denoted by $\sigma$ the coordinate
along the curve $C$. Note that each M5-brane can be thought of as
originating from the expansion of a pair of M2-branes, as in a
`standard' two-charge supertube.

Consider therefore an M5-brane with worldspace $S \times C$, where
$S$ is a complex two-surface in $\bbc{3}$ or $\bbc{4}$ and
$C$ is an arbitrary curve in $\bbe{4}$ or $\bbe{2}$,
with a worldvolume three-form flux $H=dB_2$ given by
\be
H=d\sigma \wedge \calh + dx^0 \wedge \calh' \,,
\label{H}
\ee
where again $\calh$ is understood to be pulled-back onto the
M5-brane worldvolume. $\calh'$ is a two-form determined in terms of
$\calh$ by the (generalized) self-duality condition satisfied by
$H$, but whose explicit expression will not be needed. Closure of $H$
follows from that of $\calh$. As we will see below, this $H$-flux
induces M2-brane charges on the M5-brane as in the arrays
\eqn{super3} or \eqn{super4}. We claim that this configuration preserves
1/8 or 1/16 of the supersymmetries of the M-theory Minkowski vacuum,
generated by Killing spinors $\eta$ subject to the constraints
associated to the M2-branes, that is,
\be
\Gamma_{012} \, \eta = \eta \sac
\Gamma_{034} \, \eta = \eta \sac
\Gamma_{056} \, \eta = \eta
\label{projections3}
\ee
for the three-charge supertube, and
\be
\Gamma_{012} \, \eta = \eta \sac
\Gamma_{034} \, \eta = \eta \sac
\Gamma_{056} \, \eta = \eta \sac
\Gamma_{078} \, \eta = \eta
\label{projections4}
\ee
for the four-charge supertube. Note that there is no trace of a
condition associated to the M5-branes, as expected from the absence
of M5-brane net charges. \footnote{If the curve $C$ is not closed
but instead extends to infinity, then there {\it are} net M5-brane
charges, and there is also a net linear momentum. The preserved
supersymmetries are still the ones above because the linear momentum
cancels exactly the M5-brane charges in the supersymmetry algebra
\cite{MNT}.}

To prove this, it is convenient to adopt a (static) gauge in which
$x^0$, $x^a=(x^1, \ldots, x^4)$, and $\sigma$ are worldvolume
coordinates on the M5-brane, where $\sigma$ parametrizes the
cross-section $C$, specified as $x^\mu = x^\mu(\sigma)$. For the
three-charge supertube $x^\mu = (x^7, \ldots, x^{10})$, whereas for the
four-charge supertube $x^\mu = (x^9,x^{10})$. Without loss of
generality we choose $\sigma$ to be the affine parameter along
$C$, that is,
$\delta_{\mu\nu} \, \partial_\sigma x^\mu \partial_\sigma x^\nu =1$.
The complex surface $S$ is specified as
$x^m=x^m(x^a)$, where $x^m=(x^5, x^6)$ for the three-charge supertube and
$x^m=(x^5, \ldots , x^8)$ for the four-charge supertube. In both cases,
$x^m$ satisfy the appropriate Cauchy-Riemann equations,
\be
\partial_1 x^5 = \partial_2 x^6 \sac \partial_2 x^5 = - \partial_1 x^6
\sac \ldots
\label{CR}
\ee
where the dots stand for the same expression with $\{1,2\}$ and/or
$\{5,6\}$ replaced by $\{3,4\}$ and/or $\{7,8\}$.
Note that in these coordinates the only non-zero components of
$\calh$ are
\be
\calh_{ab} = \delta_{1[a} \, \delta_{b]2} +
\delta_{3[a} \, \delta_{b]4} + \partial_{[a} x^5 \partial_{b]} x^6
\label{components3}
\ee
for the three-charge supertube, and
\be
\calh_{ab} = \delta_{1[a} \, \delta_{b]2} +
\delta_{3[a} \, \delta_{b]4} + \partial_{[a} x^5 \partial_{b]} x^6
+ \partial_{[a} x^7 \partial_{b]} x^8
\label{components4}
\ee
for the four-charge supertube. In both cases, the induced metric on the
M5-brane worldvolume takes the form
\be
ds^2 = -dx_0^2 + d\sigma^2 + g_{ab} \, dx^a \, dx^b \,,
\ee
where
\be
g_{ab} = \delta_{ab} + \partial_a x^m \, \partial_b x^n \,
\delta_{mn} \,.
\ee

The number of supersymmetries of the Minkowski vacuum preserved by
the M5-brane is the number of Killing spinors $\eta$ that satisfy
the condition $\Gamma_\rom{M5} \, \eta = \eta$ \cite{kappa}, where
$\Gamma_\rom{M5}$ is the matrix appearing in the kappa-symmetry
transformations of the M5-brane worldvolume fermions. We work with a
unit-tension M5-brane and the covariant formulation  of
\cite{PST}, which contains an auxiliary scalar field $a$ that we
eliminate by the gauge choice
$a=x^0$. Under these circumstances the kappa-symmetry matrix takes
the same form for the three- and four-charge supertubes, namely,
\be
\Gamma_\rom{M5} = \frac{\Gamma_0}{\sqrt{\det (g_{ab} + h_{ab})}} \, \left[
\frac{1}{4} \, \e^{abcd} \, \calh_{ab} \, \gamma_{cd}
- \gamma_\sigma \, \left( \calp_\sigma + \gamma_{1234} \right) \right] \,,
\label{Gamma}
\ee
where $\e^{1234} =+1$,
\be
\gamma_a = \Gamma_a + \partial_a x^m \, \Gamma_m \sac
\gamma_\sigma = \partial_\sigma x^\mu \, \Gamma_\mu
\ee
are the worldvolume Dirac matrices induced by the spacetime,
constant Dirac $\Gamma$-matrices,
$\gamma_{i_1 \ldots i_n}=\gamma_{[i_1} \gamma_{i_2} \cdots \gamma_{i_n]}$,
\be
\calp_\sigma = \frac{1}{8} \, \e^{abcd} \, \calh_{ab} \, \calh_{cd}
\label{momentum}
\ee
is the momentum density along $\partial_\sigma$, and
\be
h_{ab} \equiv \frac{1}{2} \, g_{ac} \, g_{bd} \,
\frac{\e^{cdef}}{\sqrt{\det g}} \, \calh_{ef} \,.
\ee

Making use of \eqn{projections3}, \eqn{projections4}, \eqn{CR},
\eqn{components3} and \eqn{components4}, a tedious but straightforward
calculation reveals that
\be
\gamma_{1234} \, \eta = - \calp_\sigma \, \eta \,,
\ee
where
\be
\calp_\sigma = 1 + \delta^{ab} \, \partial_a x^6 \, \partial_b x^6
\ee
for the three-charge supertube, and
\be
\calp_\sigma = 1 + \delta^{ab} \, \partial_a x^6 \, \partial_b x^6
+ \delta^{ab} \, \partial_a x^8 \, \partial_b x^8
+ \e^{abcd} \, \partial_a x^5 \, \partial_b x^6 \,
\partial_c x^7 \, \partial_d x^8
\ee
for the four-charge supertube. We note that the Cauchy-Riemann
equations imply
$\delta^{ab} \, \partial_a x^5 \, \partial_b x^5 =
\delta^{ab} \, \partial_a x^6 \, \partial_b x^6$, and analogously
with $\{5,6\}$ replaced by $\{7,8\}$.

It follows that the two terms inside the round brackets in
\eqn{Gamma} cancel each other, and hence all information about the
cross-section $C$, which is entirely encoded in $\gamma_\sigma$,
drops from the supersymmetry equation. This now reduces to
\be
\frac{1}{4} \, \e^{abcd} \, \calh_{ab} \, \Gamma_0 \gamma_{cd} \, \eta =
\sqrt{\det (g_{ab} + h_{ab})} \, \eta \,.
\ee
Using again \eqn{projections3}, \eqn{projections4}, \eqn{CR},
\eqn{components3} and \eqn{components4}, one can verify that this equation
is identically satisfied, with the determinant given by
\be
\sqrt{\det (g_{ab} + h_{ab})} =
2 \, \sqrt{\det g_{ab}} = 2 \calp_\sigma \,.
\label{det}
\ee

Essentially the same arguments given above can be used to show that
these supertube configurations saturate the bounds found in
\cite{BLW,GLWbis}. It is worth remarking, though, that the
configurations studied in detail in those references all have zero
momentum density, \ie $\calp_\sigma=\calp_a=0$.

\subsection{Physical properties}

In this section we will show that a calibrated supertube can be
regarded, at a given point, as a standard supertube with one M5
dipole and two M2 charges. For the standard supertube, supersymmetry
fixes both these densities and the shape of the tube in the
directions transverse to the cross-section (\ie it fixes
$S$) \cite{MT}.\footnote{In the simplest case
of a D2-supertube, it forces the charge densities and shape to be
either constant or those of a D2-BIon.} For a calibrated supertube,
stability is instead achieved locally in such a way that the only
restriction on this shape is that $S$ be a calibrated surface; the
charge densities may then also vary, as they are given by the
pull-back onto $S$ of the K\"ahler form. This allows the calibrated
supertube to carry, globally, more than two net charges and one
dipole, as we have seen above. In this sense, the worldvolume
three-charge supertube constructed in this paper may be regarded as
a smooth junction of three two-charge supertubes associated to the
three asymptotic regions of $S$.

The cross-section of the calibrated supertube, like that of the
standard one, is supported against collapse by the `centrifugal
force' associated to the Poynting momentum density generated by the
product of the worldvolume charge densities. To see this, we note
that the momentum density $\calp_\sigma$ is, at each point, the
product of the two M2-brane charge densities carried by the M5-brane
at that point. Recall that the M2-brane charge density in the
$ab$-directions tangent to a given point on the M5-brane is given by
the momentum density $\Pi^{ab}$ conjugate to the worldvolume
two-form potential $B_{ab}$. This is because the M5-brane action
depends only on the background-covariant combination
$H + \mathcal{A}$, where $H=dB_2$ and a pull-back onto the
M5 worldvolume of the supergravity three-form potential $A$ is
understood. It follows from this that the M2-brane charge density
carried by the M5 is
\be
\left. \frac{\partial \call_\rom{M5}}{\partial \mathcal{A}_{0ab}} \right|_{\mathcal{A}=0} =
\left. \frac{\partial \call_\rom{M5}}{\partial \dot{B}_{ab}} \right|_{\mathcal{A}=0} =
\Pi^{ab} \,.
\ee
This momentum is determined in terms of the worldspace components of
$H$ by the constraint associated to the self-duality condition
of $H$ \cite{BT}. In the present case it takes the form\footnote{For
ease of notation we are absorbing a factor of 4 in $\Pi^{ab}$ with
respect to the definition of \cite{BT}.}
\be
\Pi^{ab} = \frac{1}{2} \, \e^{abcd} \, \calh_{cd} \,,
\ee
so $\calp_\sigma$ may be rewritten as
\be
\calp_\sigma = \frac{1}{8} \, \e_{abcd} \, \Pi^{ab} \, \Pi^{cd} \,.
\ee
Now, at each point on $S$ an orthonormal basis of its tangent space
may be chosen such that the antisymmetric tensor $\Pi^{ab}$ is
skew-diagonal, with skew-eigenvalues $\Pi$ and $\Pi'$. These measure
the magnitude of the two independent M2-brane charge densities at
the given point, whereas the orthonormal basis determines their
orientations. In terms of these densities we have
\be
\calp_\sigma = \Pi \, \Pi' \,,
\label{pipi}
\ee
as anticipated.

Equation \eqn{pipi} is completely analogous to that for a two-charge
standard supertube \cite{MT,EMT}. We now show that the rest of the
relations between the charge densities, the angular momentum and the
size of the cross-section also are, at a given point, as those of
the two-charge supertube. For simplicity, we assume that $C$ is a
circle of radius $R$ in some plane, so we set $\sigma=R\,\psi$. It
is important to remember that the densities that enter these
relations are densities per unit area of $S$, obtained by
normalizing by
$\sqrt{\det g_{ab}}$ and integrating over $C$. The normalization is
most easily accounted for by working in the orthonormal basis used
to define $\Pi$ and $\Pi'$, so that $\det g_{ab} =1$. By virtue of
the second equality in \eqn{det}, this implies
\be
\calp_\sigma = \Pi \Pi' =1 \,,
\label{pone}
\ee
as for the for the standard supertube \cite{MNT}. The $C$-integrated
M2-brane densities are
\be
\varrho = \frac{1}{2\pi} \int_C d\sigma \, \Pi = R \, \Pi
\sac \varrho' = \frac{1}{2\pi} \int_C d\sigma \, \Pi' = R \, \Pi' \,,
\ee
where we have used the fact that $\sigma$ is the affine parameter
along $C$ and that $\Pi, \Pi'$ are $\sigma$-independent. It then
follows from \eqn{pone} that
\be
R = \sqrt{\varrho \varrho'} \,.
\label{R}
\ee
Similarly, the angular momentum is
\be
J_\psi = \frac{R}{2\pi}  \int_C d\sigma \, \calp_\sigma = R^2 \,,
\ee
and hence
\be
J = \varrho \varrho' \,.
\label{J}
\ee
We thus see that $R$, $J$, $\varrho$ and $\varrho'$ obey the same
relations as for a standard supertube with unit dipole, \ie
constructed from a single M5-brane. If instead $n$ M5-branes are
superposed, then these relations become
\be
R = \sqrt{\varrho \varrho'}/ n \sac J = \varrho \varrho' /n \,.
\ee

We conclude by showing that the energy of the calibrated supertube
may be written as the sum of the corresponding M2-brane charges, as
expected from supersymmetry. The M5-brane energy density $\cale$ can
be extracted from \cite{GGT}. In our case it takes the form
\be
\cale^2 = 2 \calp_\sigma^2 + \det( g + h) = 6 \calp_\sigma^2 \,.
\ee
Integrating over the M5-brane worldspace and using the definition
\eqn{momentum} of $\calp_\sigma$ we obtain the total energy
\be
E = \int_{C \times S} d\sigma \, d^4 x \,\, \cale =
\sqrt{6} \int_{C \times S} d\sigma \, d^4 x \,\, \calp_\sigma =
\frac{\sqrt{6}}{2} \int_{C\times S} d\sigma \wedge \calh \wedge \calh \,.
\ee
Employing now the definition \eqn{calh} of $\calh$, this becomes
\be
E=\frac{\sqrt{6}}{2}\, \sum_{j=1}^n\,
\int_{C\times S} d\sigma \wedge dx^{2j-1} \wedge dx^{2j} \wedge \calh =
\frac{\sqrt{6}}{2} \, \sum_{j=1}^n Q_j \,,
\ee
as we wanted to see. In the above expression appropriate pull-backs
onto the M5-brane are understood, as always, and we have used the
fact that the total charge $Q_1$ associated to an M2-brane in the
12-directions is
\be
Q_1 = \int_{C\times S} d\sigma \, d^4 x \, \Pi^{12} =
\int_{C\times S} d\sigma \, d^4 x \, \calh_{34} =
\int_{C\times S} d\sigma \wedge dx^1 \wedge dx^2 \wedge \calh \,,
\ee
and similarly for the rest of the $Q_j$. In this last equation
$\calh_{34}$ stands for the 34-component of the pull-back of
$\calh$, as opposed to the 34-component of $\calh$ itself.

\setcounter{equation}{0}
\section{Supergravity vs worldvolume description
of three-charge supertubes}
\label{sec:sugravsworld}

In the previous sections we have provided a detailed description of
three-charge supertubes within supergravity. We have also developed
a worldvolume construction of supertubes with three charges and
three dipoles in Section \ref{sec:worldv}. A third framework for
describing these systems, in terms of the microscopic CFT of D1-D5
systems, is currently under investigation. For two-charge
supertubes, the agreement between the supergravity and the
worldvolume descriptions is perfect \cite{MT,EMT,MNT}. Here we offer
some preliminary observations aimed at exploring whether a similar
connection for three-charge supertubes may exist.

In order to investigate this, let us take the two-charge supertube
as the basic `building block'. For the sake of generality and
simplicity, we phrase the discussion in terms of supergravity
charges $Q_i$, $q_i$ instead of quantized brane numbers which would
require singling out a specific U-duality frame.

For the two-charge supertube, both the worldvolume and supergravity
descriptions (and the CFT too) yield the same relations between the
parameters,
\beq
\frac{4G_5}{\pi} J_\psi=R^2 q_3=\frac{Q_1 Q_2}{q_3}\,.
\label{basic2ch}
\eeq
It is convenient to assume that the supergravity no-CCC bound is
saturated, since then the correspondence with worldvolume supertubes
is particularly simple \cite{EMT,MNT}. All the $Q_1$ and $Q_2$
branes are `dissolved' in the supertube, thus contributing to the
angular momentum, and the profile of the supertube is uniquely fixed
to be circular.

Consider now three-charge/two-dipole supertubes. Using either the
worldvolume analysis of \cite{benakraus} or our results from
supergravity one obtains \reef{qQ} (also easily interpreted within
the microscopic D1-D5 view, see \reef{32noctc1}).

Let us now regard this supertube, within the worldvolume view, as
the superposition of two two-charge supertubes of equal radius $R$,
one with parameters $(Q_1,Q_3',q_2)$, the other with
$(Q_2,Q_3'',q_1)$, and total charge $Q_3=Q_3'+Q_3''$. In terms of
the worldvolume construction of section \ref{sec:worldv}, this
simply means that the intersection between the tubes is not resolved
but remains singular. Assume that each supertube separately
satisfies corresponding relations \reef{basic2ch}. Then it is easy
to derive
\cite{benakraus}
\be
Q_3=  R^2 q_1 q_2\; \frac{Q_1 + Q_2}{Q_1 Q_2}
\,.
\label{wvQqR}\ee
The supergravity analysis, which does not allow for simple
superpositions of the two supertubes but includes instead
non-linear interactions, yields, in contrast, the CTC-bound \reef{QqR}
which, when saturated, can be written as
\be
\mathcal{Q}_3= R^2 q_1 q_2\; \frac{\mathcal{Q}_1 +
\mathcal{Q}_2}{\mathcal{Q}_1 \mathcal{Q}_2}
\,.
\label{sgQqR}\ee
(In this case $\mathcal{Q}_1=Q_1$, $\mathcal{Q}_2=Q_2$ but
$\mathcal{Q}_3<Q_3$.) This suggests
that in order to recover the supergravity expressions from the
worldvolume we must replace
\be
Q_i\to \mathcal{Q}_i\,.
\label{magic}\ee
$\mathcal{Q}_i$ is then seen to play the role of an `effective charge'.
The origin of this `replacement rule' is unclear. $Q_3$ and
$\mathcal{Q}_3$ coincide when $Q_3\gg q_1 q_2$. This is the case if we
take the limit of very large radius while keeping finite the linear
density of $Q_3$, so $Q_3\sim R$, while $q_i\sim R^0$ (see
sec.~\ref{infR}). However, at finite $R$ the worldvolume and
supergravity results differ: the supergravity radius is smaller, for
given charges. This suggests that the discrepancy might be due to
closed-string self-attraction of the ring, which would cause the tube
radius to shrink, and which would not be at work in the worldvolume
description. However, if this were the case then one might expect that,
when expressed in terms of quantized brane numbers, the string coupling
should be involved in the differences between \reef{wvQqR} and
\reef{sgQqR}, but it is not. Another possibility is that the replacement
\reef{magic} arises when the system actually forms a single supertube
(or blowing up the intersection), instead of a simple
superposition. However, the analysis of such single-brane supertubes in
Section \ref{sec:worldv} shows no evidence for this effect.
Ref.~\cite{benakraus} did also consider proper supertubes with three
charges and two dipoles, and obtained essentially \reef{wvQqR} instead
of \reef{sgQqR}.

The spin is the
sum of the spins of each supertube, so
\beq
\frac{4G_5}{\pi} J_\psi=R^2(q_1+q_2)\,
\label{jsum}\eeq
(independently of whether \reef{magic} is applied or not). This is
precisely the leading value of the angular momentum at large $R$. In
fact, it seems more appropriate to consider that the spin \reef{jsum} in
this worldvolume approach accounts exactly for the value of
$J_\psi-J_\phi$, but not for the self-dual contribution to the angular
momentum that includes $J_\phi$. In the limit $R\to\infty$ the latter
vanishes, so the discrepancy disappears.

One might remark that in the supergravity solutions $R$ is not the
proper radius of the ring (which, for $q_3 = 0$, actually diverges at
$y\to-\infty$). However, $R$, as the radius in the base space, does play
the role of the supertube radius in the two-charge supergravity
supertubes \reef{basic2ch}. In the supergravity expressions for physical
quantities $R$ can always be eliminated in favor of the physical
charges, dipoles and angular momenta (in particular, using
$J_\psi-J_\phi\propto R^2$). So we can employ it as, at least, a useful
auxiliary quantity that can be related to the worldvolume supertube
radius.

Now regard the three-charge/three-dipole supertube as the superposition
of three two-charge supertubes (see also \cite{benakraus}). Again, this
corresponds to considering that the intersection of the three M5-branes
in Section \ref{sec:worldv} is not resolved. The supertubes have parameters
$(Q_1',Q_2',q_3)$, $(Q_1'',Q_3',q_2)$, $(Q_2'',Q_3'',q_1)$. The total
charge of the $i$-th constituent is
$Q_i=Q_i'+Q_i''$. The radii of the three supertubes must be the same, so
\beq
R^2=\frac{Q_1'Q_2'}{q_3^2}=\frac{Q_1''Q_3'}{q_2^2}=\frac{Q_2''Q_3''}{q_1^2}\,.
\label{radii}\eeq
Furthermore, applying \reef{qQ} to pairwise combinations of the tubes we
obtain the constraints
\beq
q_1 Q_1'=q_3 Q_3''\,,\quad q_2Q_2''=q_1Q_1''\,,\quad q_2Q_2'=q_3Q_3'\,.
\label{pairs}
\eeq
After some algebra one can write the equation for the radius in the form
\beq
  2 \sum_{i < j} Q_i q_i Q_j q_j - \sum_i
  Q_i^2 q_i^2 = 4 R^2 q^3 \sum_i q_i\,.
\label{eqn:noctcs2}
\eeq
If we now perform the substitution \reef{magic} we recover exactly the
condition for saturation of the no-CCC
bound from supergravity, eq.~\reef{eqn:noctcs}.

The angular momentum is the
sum
\beq
\frac{4G_5}{\pi} J_\psi=
\frac{Q_1'Q_2'}{q_3}+\frac{Q_1''Q_3'}{q_2}+\frac{Q_2''Q_3''}{q_1}= R^2(q_1+q_2+q_3)
\,,
\label{spin}
\eeq
and the same commments apply as in \reef{jsum}. If we take this last
equation as giving $J_\psi-J_\phi$ instead of just $J_\psi$, then it can
be combined with \reef{eqn:noctcs2} and the substitution \reef{magic} to
reproduce the condition for saturation of \reef{eqn:noctcsb}.

We see that one key feature of the supergravity description that the
worldvolume construction does not seem to capture is the second
angular momentum $J_\phi$. We must remember that the calibrated
supertube presented in the previous section is a solution of the
{\it Abelian} theory on a single M5-brane. One may therefore
speculate that incorporating non-Abelian effects, namely working
with the theory on more than one M5-brane, is necessary to reproduce
$J_\phi$. Although this possibility cannot be discarded without
further investigation, it is hard to see how this would explain that
the solutions with a linear cross-section (\ie those obtained as the
infinite-radius limit of the ring) carry zero $J_\phi$. Here we
would like to speculate that the explanation may instead be that
$J_\phi$ is not carried by the worldvolume supertube source itself,
but that it is instead generated as a Poynting momentum by crossed
electric and magnetic {\it supergravity} gauge fields. Although the
argument we present is somewhat heuristic, it is based on general
grounds and may describe the correct physical origin of $J_\phi$. In
particular, it explains why $J_\phi=0$ for a linear cross-section.

The idea is that $J_\phi$ is given by the integral over a spacelike
hyper-surface of the $T_{0\phi}$ component of the energy-momentum
tensor that appears on the right-had side of Einstein equations.
Since the only bosonic field of $D=11$ supergravity other than the
metric is the three-form potential $\cala$, this has a unique
contribution,
\be
T_{0\phi} \sim \calf_{0mnp} \, \calf_\phi^{\;\;mnp}
\,,
\ee
which for our solution takes the form
\be
T_{0\phi} \sim \calf_{012m} \, \calf_\phi^{\;\;12m} +
\calf_{034m} \, \calf_\phi^{\;\;34m} +
\calf_{056m} \, \calf_\phi^{\;\;56m} \,.
\label{con}
\ee
This is indeed a product of electric and magnetic components of
$\calf$. We wish to argue on general grounds that this must be
non-zero for a brane array as \eqn{intersection} except if the
$\psi$-direction is a straight line.

Indeed, we know the first M2-brane must generate a non-zero
component $\cala_{012}$, whose magnitude must be proportional to
$Q_1$. Similarly, the first M5-brane must source a non-zero
component $\tilde{\cala}_{03456\psi}$ of the six-form potential dual
to $\cala$, whose magnitude must be proportional to $q_1$. Analogous
statements apply to the other two M2/M5 pairs, so let us concentrate
on the first pair.

The key difference between a linear and circular (or, more
generally, any non-linear) cross-section is that, in the linear
case, all fields may depend on a single radial coordinate
$\rho$ in $\bbe{4}$ (see \eqn{rhoThetabase}), because of
$SO(3)$ rotational symmetry around the string. This means that the
non-zero components of the gauge potentials above lead to the
non-zero components $\calf_{012\rho}, \tilde{\calf}_{03456\psi\rho}$
of the corresponding field strengths. Hodge-dualizing
$\tilde{\calf}$ we see that the only magnetic component of
$\calf$ is $\calf_{\phi 12 \Theta}$. It follows that the
contractions in \eqn{con} vanish and hence $J_\phi=0$. This is in
fact the only result compatible with $SO(3)$ symmetry, since any
non-zero angular momentum in the $\bbe{3}$ space transverse to the
string would break this symmetry down to $U(1)$. In the case of a
circular cross-section this breaking is already present from the
beginning by the choice of plane in which the ring lies.

Indeed, for a circular cross-section (in fact, for {\it any}
non-linear cross-section) the contractions \eqn{con} do not vanish.
This is because now the fields depend on two coordinates
\be
\rho_1 = \rho \cos \Theta \sac \rho_2 = \rho \sin \Theta \,,
\ee
so $\tilde{\calf}$ has non-zero components
$\tilde{\calf}_{03456\psi\rho_1}$ and $\tilde{\calf}_{03456\psi\rho_2}$.
Dualizing we find that $\calf$ has non-zero magnetic components
$\calf_{\phi 12 \rho_1}$ and $\calf_{\phi 12 \rho_2}$ and therefore
that the contractions in \eqn{con} do not vanish in general.
Moreover, the electric components are proportional to $Q_i$, whereas the
magnetic ones are proportional to $q_i$, so naively $T_{0\phi}$ is
proportional to $Q_1 q_1 + Q_2 q_2 + Q_3 q_3$. This gives the first
term in $J_\phi$, and therefore it satisfies the requirement that it
be zero for a standard two-charge/one-dipole supertube. However, it
misses the second term in $J_\phi$, proportional to
$q_1 q_2 q_3$. This is because our argument ignored the fact that
the electric component $\cala_{012}$ is not just proportional to
$Q_1$ but actually contains a term proportional to $q_2 q_3$ (see the
expression
\eqn{eqn:Hi} for $H_1$). Analogously, the electric components
$\cala_{034}$ and $\cala_{056}$ contain terms proportional to
$q_1 q_3$ and $q_1 q_2$, respectively. Each of these electric
components, when multiplied by the corresponding magnetic component,
gives a term proportional to $q_1 q_2 q_3$.

The argument above suggests that, although the precise value of
$J_\phi$ depends on some details of the solution, the fact that
it is non-vanishing follows on general grounds from the presence of
brane sources oriented as in the \eqn{intersection}. From a
mechanical viewpoint, one may say that in order to bend the first
M5-brane to close one of its directions into a circle, in the
presence of the first M2-brane, an angular momentum must be
generated by the crossed electric and magnetic fields they source.

Although tentative, the observations in this section point to
non-trivial connections between the worldvolume and supergravity
descriptions of supertubes with three charges. The justification of
\reef{magic} remains an important open issue. It is presumably
significant that it makes appearance only when the BPS equations
solved by the supergravity solution are non-linear. The perfect
agreement observed between worldvolume and supergravity for
two-charge supertubes would then seem to be a chance effect of the
linearity of the system. In fact there does not seem to be any {\it
a priori} reason to expect perfect agreement. Supersymmetry, in
particular, does not provide any clear reason for this. The physical
origin of $J_\phi$ needs further investigation too.

Finally, one might note that the supergravity constraints, derived
by requiring absence of causal anomalies, do not actually fix the
angular momentum, for given charges, but instead impose an upper
bound on it. The cases where the bound is not saturated include the
black supertubes with non-zero area. From the worldvolume
perspective, it has been argued that two-charge supertubes with
profiles other than circular, which do not saturate the bound on the
angular momentum \cite{MNT}, are degenerate. Upon quantization,
their degeneracy is equal to the degeneracy of the Ramond ground
states of the supersymmetric D1-D5 string
\cite{mathur1,othertubes}. Thus it would seem natural to conjecture
that the degeneracy of three-charge worldvolume supertubes obtained
by quantizing their moduli space of arbitrary profiles, can
similarly reproduce the entropy of three-charge black supertubes,
possibly after making the substitution \reef{magic}. If we consider
such a supertube as made of three superposed two-charge supertubes
as in the construction above, it is easy to see that the no-CCC
bound comes out correctly --- after making the replacement
\reef{magic} --- but the entropy is too small.
Hence supertubes with $L>0$ are, not surprisingly, quite more
complicated than these simple composites. Even if one considered the
worldvolume three-charge resolved supertubes constructed in the
previous section, it might still be that a calculation of their
entropy fails to reproduce exactly the Bekenstein-Hawking entropy of
three-charge supertubes. Instead, our analysis suggests that the
worldvolume description might only reproduce the supergravity
results up to the replacement \reef{magic}, and then accounting only
for the non-self-dual part of the angular momentum, $J_\psi-J_\phi$.

\setcounter{equation}{0}
\section{Concluding remarks}
\label{sec:concl}

Supersymmetric five-dimensional black holes are of considerable interest
in string theory because they admit a simple microscopic description in
terms of D-branes \cite{stva}. Until now, the largest known family of
such black holes was the four-parameter BMPV family \cite{bmpv}. Our
work has shown that this is a limiting case of a larger seven-parameter
family of supersymmetric black rings. Clearly the challenge now is to
obtain a microscopic description of these solutions that correctly
accounts for their entropy.

It is natural to ask whether we have now exhausted the catalogue of
supersymmetric $D=5$ black holes. One might be tempted to speculate that
there are many further surprises to be discovered. These are
constrained, however, by the analysis of possible near-horizon
geometries of supersymmetric black holes in
Refs.~\cite{reall:02,gutowski:04}, which allows for only three
possibilities (at least for the class of $D=5$ supergravity theories
considered in appendix \ref{app:U1N}): (i) flat space, (ii) $AdS_3
\times S^2$, and (iii) near-horizon BMPV, with corresponding horizon
geometry (i) $T^3$, (ii) $S^1 \times S^2$ or (iii) (possibly a quotient
of) a homogeneously squashed $S^3$. It was also shown that the only
asymptotically flat solution of type (iii) is the BMPV black hole. Hence
if there exist any further supersymmetric black hole solutions then they
must either have a flat near-horizon geometry or they must be black
rings distinct from the ones presented here.

Could there exist supersymmetric black rings distinct from the ones presented here?
Our worldvolume analysis of three-charge supertubes suggests that
there might exist corresponding supergravity solutions with profiles
other than circular. If such solutions had horizons then
the results of \cite{reall:02,gutowski:04}  prove that the near-horizon geometry must be
the same as for the circular black rings presented here, so deviations from circularity would only be apparent away from the horizon. On the other hand, refs.~\cite{mathur2,lunin,sm} have constructed regular horizon-free solutions with three charges and suggest the existence of a larger
class of them. So maybe solutions with non-circular cross-sections
would belong to this class instead. Clearly, the space of physically
relevant D1-D5-P solutions is far from being completely mapped out.

The fact that three-charge supersymmetric black rings exhibit
non-uniqueness might seem at first to be a difficulty for a
microscopic description. However, what appears to be an obstacle may
actually be a very useful ingredient towards a more complete
understanding of the D1-D5-P system. It has been argued in
\cite{EE,RE}, in the context of near-extremal solutions, that string
theory could contain the necessary states to account for black
rings. One needs to appropriately identify the dipole constituents,
or the phase in which the strings are. From a thermodynamical
viewpoint, solutions that are characterized by the same asymptotic
charges should be regarded as being only locally stable in general.
If we maximize the entropy by varying the two independent dipoles we
find a {\it unique} solution, which should be the only globally
thermodynamically stable configuration. Still, it would seem that
all local equilibrium states, and not only the global maxima, should
admit a microscopic description. It would indeed be very surprising
if string theory could not account for solutions of its low energy
supergravity limit that seem completely pathology-free.

A remarkable aspect of supergravity solutions is the way in which
they capture highly non-trivial constraints between parameters that
arise in a microscopic description. It was already known for the
BMPV black hole and for the two-charge supertube that the condition
that CCCs be absent yields the correct upper bounds on angular
momentum required by the microscopic CFT or worldvolume theory.\footnote{
Such constraints from supergravity do also arise,
alternatively, from the requirement that the solution admits a
thermal deformation.} For supertubes with three charges we have
observed similar non-trivial results in section
\ref{sec:sugravsworld}. However, the worldvolume description falls
just short of perfect agreement with supergravity: a basis has to be
found for the simple (partial) fix of \reef{magic}. The origin of
$J_\phi$ also needs to be better understood. Although we have presented a
speculative explanation for this origin, further investigation is
clearly needed to establish a connection, at the same level as that
for two-charge supertubes, between the worldvolume and supergravity
constructions of three-charge supertubes studied in this paper.

\section*{Acknowledgments}

\noindent
We thank J.~Gauntlett, G.~Horowitz,  D.~Marolf and R.~Myers
for useful discussions. This work was presented by HE and HSR at the
GR-17 conference in Dublin, July 18-23, 2004. We would like to thank
the audience, in particular V.~Hubeny, M.~Rangamani and S.~Ross, for
positive feedback. HE was supported by the Danish Research Agency and
NSF grant PHY-0070895. RE was
supported in part by UPV00172.310-14497, FPA2001-3598, DURSI
2001-SGR-00188, HPRN-CT-2000-00131. DM is supported in part by funds
from NSERC of Canada. HSR was supported in part by the National
Science Foundation under Grant No.~PHY99-07949.

\vfill

\newpage

\appendix

\section*{Appendices}

\setcounter{equation}{0}
\section{Derivation of the ring in minimal 5D supergravity}
\label{app:minsugra}

\subsection{Supersymmetric solutions of minimal supergravity}
\label{sec:minsoln}

Minimal $D=5$ supergravity is a theory with eight supercharges with bosonic action
\be
I = \frac{1}{16 \pi G_5} \int \left( R \star_5 1 - 2 F \wedge \star_5 F
- \frac{8}{3\sqrt{3}} F \wedge F \wedge A \right),
\ee
where $F=dA$. Any supersymmetric solution of this theory admits a
globally defined non-spacelike Killing vector field $V$
\cite{gibbons:93} that cannot vanish \cite{reall:02}. In a region where
$V$ is timelike, coordinates $(t,x^m)$ can be introduced so that $V =
\partial/\partial t$ and the line element can be written
\be
\label{eqn:metric}
 ds^2 = -f^2 (dt + \omega)^2 + f^{-1} h_{mn} dx^m dx^n,
\ee
where $h_{mn}$ is a Riemannian metric on a four-dimensional space
referred to as the ``base space" ${\cal B}$. The metric $h_{mn}$, scalar
$f$ and 1-form $\omega \equiv \omega_m dx^m$ are all independent of $t$.
Supersymmetry implies that $h_{mn}$ is a hyper-K\"ahler metric on ${\cal
B}$ and that the
Maxwell field strength is given by \cite{harveyetal}
\be
 F = \frac{\sqrt{3}}{2} d \left[f(dt + \omega) \right] - \frac{1}{\sqrt{3}} G^+,
\label{eqn:field}\ee
with
\be
 G^+ \equiv \frac{1}{2} f \left(d\omega + \star_4 d\omega \right),
\label{eqn:Gplus}\ee
where $\star_4$ denotes the Hodge dual on ${\cal B}$ with respect to the
metric $h_{mn}$ with orientation defined so that the complex structures
are anti-self dual. These conditions are necessary for supersymmetry; it
turns out that they are also sufficient. In the orthonormal basis $e^0 =
f(dt+\omega)$, $e^i = f^{-1/2} \hat{e}^i$ with $\hat{e}^i$ an
orthonormal basis for $h_{mn}$, the Killing spinor equation is solved by
\cite{harveyetal}
\be
\label{eqn:spinor}
 \epsilon(t,x) = f^{1/2} \eta(x),
\ee
where $\eta$ is any chiral spinor on ${\cal B}$ that is covariantly
constant with respect to the Levi-Civita connection of $h_{mn}$. This
implies that any supersymmetric background preserves at least $1/2$
supersymmetry. In fact the only allowed fractions of supersymmetry are
$0,1/2$ and $1$. This is easy to understand by noting the isomorphism
$\rom{Spin}(1,4) = \rom{Sp}(1,1)$ under which the irreducible spinor representation
becomes a quaternion doublet \cite{bryant:00}. The Killing
spinor equation is linear so any solution can be multiplied by a constant
quaternion hence the general solution must have a multiple of 4 real degrees of
freedom.

We are interested in supersymmetric {\it solutions} so we must also
impose the equations of motion for this theory. The Bianchi identity for
$F$ gives
\be
\label{eqn:bianchi}
 dG^+ = 0,
\ee
and the Maxwell equation reduces to \cite{harveyetal}
\be
\label{eqn:maxwell}
\nabla^2 f^{-1} = \frac{4}{9} \left( G^+ \right)^2 \equiv \frac{2}{9}
G^+_{mn} G^{+mn},
\ee
where $\nabla^2$ is the Laplacian on ${\cal B}$ with respect to $h$.
The Einstein equation is automatically satisfied as a consequence of the
above equations \cite{harveyetal}.

\subsection{Minimal supersymmetric black rings}

We start by choosing the base space to be flat space written in the form
of equation \eqn{eqn:flatspace} with orientation $\epsilon_{y\psi
x\phi} = +1$ (note that flat space admits anti-self-dual hyper-K\"ahler
structures of either orientation).
We make the Ansatz
\be
 \omega = \omega_\phi(x,y) d\phi + \omega_\psi (x,y) d\psi.
\ee
Equation \eqn{eqn:bianchi} gives
\ba
\label{eqn:bianchi2}
\partial_x\left[ f(\omega_{\psi,y}+\omega_{\phi,x})\right]
&=& \partial_y\left[
f\left(\omega_{\psi,x}-\frac{y^2-1}{1-x^2}\omega_{\phi,y} \right)
\right]\,, \nn
\partial_y \left[f(\omega_{\psi,y}+\omega_{\phi,x}) \right]
&=& \partial_x \left[f\left(
\omega_{\phi,y}-\frac{1-x^2}{y^2-1}\omega_{\psi,x}\right)\right]\,.
\ea
We assume
\be
  \label{simpleGplus}
  \omega_{\psi , x} = \frac{y^2-1}{1-x^2} \, \omega_{\phi , y}\,,
\ee
so \eqn{eqn:bianchi2} reduces to
\be
\label{eqn:bianchi3}
 f(\omega_{\psi,y}+ \omega_{\phi,x})=\frac{3}{2}  q
\label{fomega}
\ee
where $q$ is a constant. This determines
\be
G^+ = \frac{3}{4} q (dx\wedge d\phi+dy\wedge d\psi)
\ee
and
\be
(G^+)^2=\frac{9 q^2 (x-y)^4}{8 R^4}\,.
\label{Gplus2}
\ee
Now we seek a solution to equation \eqn{eqn:maxwell}. We obtain it
as a harmonic piece from solutions to the homogeneous
Laplace equation $\nabla^2 f^{-1}=0$, plus a solution to the
inhomogeneous Poisson equation
sourced by \reef{Gplus2}. It seems reasonable to take the harmonic piece
to contain a term $\propto x-y$, since this is the solution to the
Laplace equation in $\bbr{4}$ with delta-function sources on a
circle of radius $R$, which appears in two-charge supertube solutions
\cite{HE,EE}. In order to solve the Poisson equation, one looks
for a simple enough function of $x$ and $y$, which is antisymmetric
under $x\leftrightarrow y$, vanishes at infinity ($x,y \rightarrow
-1$), and is singular on the circle at
$y\to-\infty$. A quick survey leads to the solution
$-q^2(x^2-y^2)/(4R^2)$. Then we take
\be
 f^{-1} = 1 + \frac{Q-q^2}{2R^2} (x-y) - \frac{q^2}{4R^2} (x^2 - y^2),
\ee
where we have normalized so that $f \rightarrow 1$ as $x,y \rightarrow
-1$. Higher harmonics in $f^{-1}$, such as $xy(x-y)$, would lead
to more singular behavior on the ring and so are discarded.

It remains to solve equations \eqn{simpleGplus} and \eqn{eqn:bianchi3}
to determine $\omega$. Equation \eqn{simpleGplus} is equivalent to
\be
 \omega_\phi = (1-x^2) \partial_x W, \qquad \omega_\psi = (y^2-1) \partial_y W,
\ee
for some function $W(x,y)$. Substituting this into equation
\eqn{eqn:bianchi3} gives
\be
\label{eqn:Feq}
\partial_x \left( (1-x^2) \partial_x W \right) + \partial_y \left( (y^2
-1) \partial_y W \right) = \frac{3}{2} q f^{-1}.
\ee
We demand that $\omega_\phi$ vanish at $x = \pm 1$ and $\omega_\psi$
vanish at $y = -1$, hence $W$ must be finite at $x= \pm 1$ and at $y
= -1$. Looking for a solution of the form $W = X_1(x) + Y_1(y)$ leads to
\ba
 X_1'(x) &=& -\frac{q}{8R^2} \left(3Q - q^2(3+ x) \right),\nn
 Y_1'(y) &=& -\frac{3q}{2(1-y)} - \frac{q}{8R^2} \left(3Q - q^2(3 + y) \right).
\ea
We are free to add a solution of the homogeneous equation (i.e.
equation \eqn{eqn:Feq} with $q=0$). If we look for solutions of the form
$X_2(x)Y_2(y)$ subject to the above regularity conditions then we are
led to $X_2(x) Y_2(y) \propto P_l(x)P_l(y)$ where $P_l$ are Legendre
polynomials. In general, we can add an infinite sum of such terms to
$W$. However,
in order to avoid the orbits of $\partial/\partial \phi$ and
$\partial/\partial \psi$ being closed timelike curves as $y \rightarrow
-\infty$, $\omega_\phi$ and $\omega_\psi$ can diverge no faster than
$y^2$, which restricts us to $l \le 2$. More careful inspection reveals
that the norm of $\partial/\partial \psi$ diverges as $y \rightarrow -
\infty$ if a $l=2$ term is present so we need $l \le 1$, corresponding
to the solution
\ba
\label{omegask}
 \omega_\phi &=& - \frac{q}{8R^2} (1-x^2) \left[3Q - q^2 ( 3 +
 x ) + ky \right] \nonumber \\
 \omega_\psi &=& \frac{3q}{2} (1+y)  - \frac{q}{8R^2}(y^2-1)
 \left[3Q - q^2 (3 + y) + kx \right],
\ea
where $k$ is a constant. We then find, near $y=-\infty$,
\beq
g_{\psi\psi}= \frac{1}{2} (k + q^2 ) x y +{\cal O}\left(y^0 \right)\,,
\eeq
so we must choose $k=-q^2$ to prevent some orbits of $\partial/\partial
\psi$ from being closed timelike curves (CTCs). This completes the
derivation of the solution given in \cite{EEMR}.
\setcounter{equation}{0}
\section{Supersymmetric black rings in $U(1)^N$ theories}
\label{app:U1N}

\subsection{The theory}

The method of \cite{harveyetal} has been generalized to the case of
minimal supergravity coupled to $N-1$ abelian vector multiplets
with scalars taking values in a symmetric space \cite{gutowski:04a,
gutowski:04b}. The action for such a theory is
\be
\label{eqn:5daction}
 I = \frac{1}{16 \pi G_5} \int \left( R \star 1 - G_{IJ} dX^I \wedge
\star dX^J - G_{IJ} F^I \wedge \star F^J - \frac{1}{6} C_{IJK} F^I
\wedge F^K \wedge A^K \right) \,,
\label{5action}
\ee
where $I,J,K=1,\ldots,N$. The constants $C_{IJK}$ are symmetric in
$(IJK)$ and obey
\be
 C_{IJK} C_{J' (LM}C_{PQ)K'} \delta^{JJ'} \delta^{KK'} = \frac{4}{3}
\delta_{I(L} C_{MPQ)}.
\ee
The $N-1$ dimensional scalar manifold is conveniently parametrized by
the $N$ scalars
$X^I$, which obey the constraint\footnote{Given this constraint, the
  scalars should be written in terms of unconstrained variables before
  the action is varied.}
\be
 \frac{1}{6} C_{IJK} X^I X^J X^K = 1.
\ee
It is then convenient to define
\be
 X_I \equiv \frac{1}{6} C_{IJK} X^J X^K,
\ee
so $X_I X^I = 1$. The matrix $G_{IJ}$ is defined by
\be
 G_{IJ} = \frac{9}{2} X_I X_J - \frac{1}{2} C_{IJK} X^K,
\ee
with inverse
\be
 G^{IJ} = 2 X^I X^J - 6 C^{IJK} X_K,
\ee
where $C^{IJK} \equiv C_{IJK}$. We also have
\be
 X^I = \frac{9}{2} C^{IJK} X_J X_K.
\ee
Reducing the $D=11$ supergravity action,
\be
\label{eqn:11daction}
I = \frac{1}{16 \pi G_{11}} \int \left( R_{11} \star_{11} 1 -
\frac{1}{2} {\cal F} \wedge \star_{11}{\cal F} - \frac{1}{6} {\cal F}
\wedge {\cal F} \wedge {\cal A}
\right) \,,
\ee
to $D=5$ on $T^6$ using the Ansatz \eqn{11sol} with the constraint
\eqn{eqn:constr}
yields precisely the action \eqn{5action} with $N=3$, $C_{IJK}=1$ if $(IJK)$ is a
permutation of $(123)$ and $C_{IJK}=0$ otherwise, and
\be
 G_{IJ} = \frac{1}{2} {\rm diag} \left( (X^1)^{-2},(X^2)^{-2},(X^3)^{-2} \right).
\ee

\subsection{Supersymmetric solutions}

Any supersymmetric solution of this theory must admit a non-spacelike
Killing vector field $V$ \cite{gutowski:04a} so, in a region where $V$
is timelike, we can introduce coordinates just as in the minimal theory
described in Appendix \ref{app:minsugra}.
Supersymmetry again implies that $({\cal B},h)$ is a hyper-K\"ahler
manifold and that the
Maxwell fields can be written \cite{gutowski:04a}\footnote{
Set $\chi=0$ in \cite{gutowski:04a} to obtain these results.}
\be
 F^I = d \left(f X^I (dt + \omega) \right) + \Theta^I,
\ee
where $\Theta^I$ are self-dual $2$-forms on ${\cal B}$ satisfying
\be
\label{eqn:gpsol}
 X_I \Theta^I = -\frac{2}{3} G^+.
\ee
The above conditions are both necessary and
sufficient for the existence of a supercovariantly constant
spinor of the same form \eqn{eqn:spinor} as in the minimal theory
\cite{gutowski:04b}.

We also need to satisfy the
equations of motion. The Bianchi identity for $F^I$ is
\be
 d\Theta^I = 0,
\ee
and the Maxwell equation is \cite{gutowski:04a}
\be
\label{eqn:maxwell3}
 \nabla^2 \left( f^{-1} X_I \right) = \frac{1}{6} C_{IJK} \Theta^J
\cdot \Theta^K,
\ee
where, $\nabla^2$ is the Laplacian on ${\cal B}$ and,
for $2$-forms $\alpha$ and $\beta$ on ${\cal B}$,
$\alpha \cdot \beta \equiv (1/2) \alpha^{mn} \beta_{mn}$, raising
indices with $h^{mn}$. The remaining equations of motion are satisfied
automatically \cite{gutowski:04b}.

\subsection{Supersymmetric black rings}

We now want to generalize our black ring solution of the minimal
theory to a solution of the theory \reef{eqn:5daction}. We proceed by
analogy with
the minimal theory. First we choose ${\cal B}$ to be flat space
written in the form \reef{eqn:flatspace}. Next we need to find some closed,
self-dual 2-forms $\Theta^I$ on ${\cal B}$. We already know one
example of such a 2-form from the minimal theory, namely $G^+$. This
suggests the Ansatz
\be
 \Theta^I = -\frac{1}{2} q^I \left( dy \wedge d\psi + dx \wedge
 d\phi \right),
\ee
for some constants $q^I$.
Equation \reef{eqn:maxwell3} reduces to
\be
 \nabla^2 \left(f^{-1} X_I \right) = \frac{1}{12 R^4} C_{IJK} q^J q^K
 (x-y)^4.
\ee
Comparison with the corresponding equation of the minimal theory
immediately provides a solution:
\be
\frac{1}{3} H_I \equiv f^{-1} X_I = \bar{X}_I + \frac{1}{6R^2} \left(
Q_I - \frac{1}{2}
 C_{IJK} q^J q^K \right) (x-y) - \frac{1}{24 R^2} C_{IJK}
 q^J q^K  (x^2-y^2),
\ee
where the constants $Q_I$ are arbitrary but demanding $f \rightarrow
1$ at infinity implies that
the constants $\bar{X}_I$ must obey the same algebraic restrictions
as $X_I$. These restrictions also imply
\be
 f^{-3} = \frac{1}{6} C^{IJK} H_I H_J H_K.
\ee
Next, from equation \reef{eqn:gpsol} we have
\be
 G^+ = - \frac{3}{2} X_I \Theta^I,
\ee
which we have to solve to determine $\omega$. A natural Ansatz is
\be
 \omega = q^I \omega_I
\ee
where $\omega_I$ obeys
\be
 \frac{1}{2} \left( d\omega_I + \star_4 d\omega_I\right) = \frac{1}{4}
 H_I \left(dy \wedge d\psi + dx \wedge d\phi \right).
\ee
This is exactly the same as the equation we had to solve in the minimal
theory and can be solved in the same way -- we have
\be
\omega_{I\phi} = (1-x^2) \partial_x W_I, \qquad \omega_{I\psi} = (y^2-1)
\partial_y W_I
\ee
where $W_I$ is regular at $x = \pm 1$ and $y=-1$ but can diverge
logarithmically at $y=1$, and must obey
\be
 \partial_x \left( (1-x^2) \partial_x W_I \right) + \partial_y \left(
 (y^2-1) \partial_y W_I \right) = \frac{1}{2} H_I.
\ee
We can just read off a solution by carrying over results from
the minimal theory:
\be
 \omega_{I\phi} = - \frac{1}{8 R^2} (1-x^2) \left[ Q_I -
   \frac{1}{6 R^2} C_{IJK} q^J q^K \left( 3 + x + y
   \right) \right],
\ee
\be
 \omega_{I \psi} = \frac{3}{2} (1+y) \bar{X}_I - \frac{1}{8R^2}
(y^2-1) \left[
   Q_I - \frac{1}{6 R^2} C_{IJK} q^J q^K \left( 3 + x + y
   \right) \right].
\ee
So finally we have
\be
 \omega_\phi = - \frac{1}{8 R^2} (1-x^2) \left[ q^I Q_I -
   q^3 \left( 3 + x + y
   \right) \right] ,
\ee
\be
 \omega_\psi = \frac{3}{2} (1+y) q^I \bar{X}_I - \frac{1}{8R^2}
(y^2-1) \left[
   q^I Q_I - q^3 \left( 3 + x + y \right) \right],
\ee
where
\be
 q^3 \equiv \frac{1}{6} C_{IJK} q^I q^J q^K.
\ee
The electric charges are given by
\be
 \mathbf{Q}_I = \frac{1}{8\pi G_5} \int G_{IJ} \star F^J = \frac{
\pi}{4G_5} Q_I.
\ee
The mass and angular momenta can be read off by comparing the
asymptotics of the above solution with the minimal ring. We find
\be
 M = \bar{X}^I \mathbf{Q}_I = \frac{\pi}{4G_5} \bar{X}^I Q_I.
\ee
\be
 J_{\phi} = \frac{\pi}{8G_5} \left(  q^I Q_I - q^3 \right),
\qquad J_{\psi} = \frac{\pi}{8G_5} \left( 6R^2 q^I \bar{X}_I +
q^I Q_I - q^3 \right).
\ee
The solutions of section \ref{sec:gensol} are obtained by replacing indices
$I,J,K$ with $i,j,k$, choosing $\bar{X}_i = 1/3$, i.e., $\bar{X}^i =
1$, and defining $q_i = q^i$.

\setcounter{equation}{0}
\section{Positivity of the $\phi$-$\psi$ metric}
\label{app:noctcs}

The determinant of the $\phi$-$\psi$ part of the five-dimensional metric
\reef{eqn:5dsol} is
\be
 \Delta \equiv f^{-2} h_{\phi\phi} h_{\psi \psi} - f \omega_\phi^2
 h_{\psi \psi} - f \omega_\psi^2 h_{\phi \phi}\,,
\ee
where $h_{mn}$ is the base space metric \reef{eqn:flatspace}.
If $\Delta$ is positive then the $\phi$-$\psi$ metric is either positive
definite or negative definite. To see which, note that
\be
 \Delta = f^{-1} h_{\phi \phi} g_{\psi\psi} - f \omega_{\phi}^2
h_{\psi \psi},
\ee
so $\Delta >0$ implies $g_{\psi \psi}>0$. Hence the $\phi$-$\psi$ metric
is positive definite when $\Delta >0$.

We find that
\ba
\label{eqn:delta}
 \frac{(x-y)^4}{(1-x^2) (-1-y) f R^4} \Delta &=& \frac{
(y^2-x^2)(1-y)(x-y)^2}{64R^6} \left( 2 \sum_{i < j} \mathcal{Q}_i q_i
\mathcal{Q}_j q_j - \sum_i
\mathcal{Q}_i^2 q_i^2 - 4 R^2 q^3 \sum_i q_i \right) \nn
 &+& \frac{ (x-y)^2 (1-y)}{8R^4} \left[X - (x+y) q^3 \sum_i q_i +
(1-x) \sum_{i \ne j} \mathcal{Q}_i q_i q_j \right] \\ &+& \frac{(x-y)^2}{4R^2} Y
+ \frac{(1-y)}{4R^2} \left[ (y^2-x^2) \sum_{i < j} q_i q_j +
2(x-y) \sum_i \mathcal{Q}_i \right] + (1-y) \nonumber
\ea
where $\mathcal{Q}_i$ is defined in equation \reef{eqn:adef} and
\be
X \equiv \frac{\mathcal{Q}_1 \mathcal{Q}_2 \mathcal{Q}_3}{R^2} (x-y) +
(1+y) \sum_i \mathcal{Q}_i q_i^2\,,
\quad Y \equiv \frac{(1-y)}{R^2} \sum_{i < j} \mathcal{Q}_i \mathcal{Q}_j + (1+y)
\left( \sum_i q_i \right)^2.
\ee
We derived this expression by first grouping together terms involving the
same power of $R$. For $\Delta$ to be positive as $y \rightarrow -
\infty$ we need
\be
  2 \sum_{i < j} \mathcal{Q}_i q_i \mathcal{Q}_j q_j - \sum_i
  \mathcal{Q}_i^2 q_i^2 \ge 4 R^2 q^3 \sum_i q_i.
\label{eqn:ctcineq}
\ee
For $q_i>0$, this inequality implies that the $\mathcal{Q}_i$ must lie
in the region interior to
one sheet of a double sheeted hyperboloid. It is easy to see that it
cannot be satisfied if any of the $\mathcal{Q}_i$ vanishes hence the allowed
region lies entirely within the positive octant\footnote{
The other sheet lies in the negative octant but shall we disregard this region.}
of $R^3$, i.e., $\mathcal{Q}_i>0$.

All of the remaining terms in $\Delta$ are manifestly positive except
for $X$ and $Y$. These can be seen to be positive as follows:
\ba
 Y &>& -(1+y) \left[ \frac{1}{R^2} \sum_{i < j} \mathcal{Q}_i \mathcal{Q}_j - \left(
\sum_i q_i \right)^2 \right]
 = - \frac{(1+y)}{4 R^2 q^3} \left[4 q^3  \sum_{i < j} \mathcal{Q}_i \mathcal{Q}_j - 4
R^2 q^3 \left( \sum_i q_i \right)^2 \right] \nn
&\ge& - \frac{(1+y)}{4 R^2 q^3} \left[4 q^3 \sum_{i < j} \mathcal{Q}_i
\mathcal{Q}_j - \left( \sum_i q_i \right) \left( 2 \sum_{i < j}
\mathcal{Q}_i q_i \mathcal{Q}_j q_j - \sum_i \mathcal{Q}_i^2 q_i^2
\right) \right]\nn
&=& - \frac{(1+y)}{4 R^2 q^3} \left[ q_1 \left( \mathcal{Q}_2 q_2 +
\mathcal{Q}_3 q_3 - \mathcal{Q}_1 q_1 \right)^2 + q_2 \left(
\mathcal{Q}_3 q_3 + \mathcal{Q}_1 q_1 - \mathcal{Q}_2 q_2 \right)^2 +
q_3 \left( \mathcal{Q}_1 q_1 + \mathcal{Q}_2 q_2 - \mathcal{Q}_3 q_3
\right)^2 \right] \nn
&\ge& 0 \nonumber .
 \ea
Here the first inequality is just $1-y > -1-y$ and the second follows
from \reef{eqn:ctcineq}. For $X$ we have
\ba
X &\ge& -(1+y) \left[ \frac{\mathcal{Q}_1 \mathcal{Q}_2
\mathcal{Q}_3}{R^2} - \sum_i \mathcal{Q}_i q_i^2
\right] \nn
&=& -\frac{(1+y)}{4 R^2 q^3 \sum_i q_i} \left[ 4 \mathcal{Q}_1
\mathcal{Q}_2 \mathcal{Q}_3 q^3 \sum_i q_i - 4 R^2 q^3 \left( \sum_i q_i
\right) \left(\sum_j \mathcal{Q}_j
q_j^2 \right) \right] \nn
&\ge& -\frac{(1+y)}{4 R^2 q^3 \sum_i q_i} \left[ 4 \mathcal{Q}_1
\mathcal{Q}_2 \mathcal{Q}_3 q^3
\sum_i q_i - \left(\sum_j \mathcal{Q}_j
q_j^2 \right) \left( 2 \sum_{i < j} \mathcal{Q}_i q_i \mathcal{Q}_j q_j - \sum_i
  \mathcal{Q}_i^2 q_i^2 \right) \right] \nn
&=& -\frac{(1+y)}{4 R^2 q^3 \sum_i q_i} \left[\mathcal{Q}_1 q_1^2 \left(
\mathcal{Q}_2 q_2 + \mathcal{Q}_3 q_3 - \mathcal{Q}_1 q_1 \right)^2 +
\mathcal{Q}_2 q_2^2 \left( \mathcal{Q}_3 q_3 + \mathcal{Q}_1 q_1 -
\mathcal{Q}_2 q_2
\right)^2 \right. \nn
 && +\left.
\mathcal{Q}_3 q_3^2 \left( \mathcal{Q}_1 q_1 + \mathcal{Q}_2 q_2 -
\mathcal{Q}_3 q_3 \right)^2
\right] \nn
&\ge& 0 \nonumber .
\ea
The first inequality is just $x-y \ge -1-y$ and the second follows
from \reef{eqn:ctcineq}.
In summary, the only condition required for $\Delta$ to be positive is
\reef{eqn:ctcineq}.

\setcounter{equation}{0}
\section{Extension through the horizon}
\label{app:through}

\subsection{Five dimensional solution}

The solution \reef{eqn:5dsol} can be extended through $y = -\infty$ in the
same way as the minimal ring \cite{EEMR}: let $\bar{r} = -R/y$ and
\be
dt = dv - A(\bar{r}) d\bar{r}, \qquad d\phi = d\phi' - B(\bar{r})
d\bar{r}, \qquad d\psi =
d\psi' -B(\bar{r}) d\bar{r},
\label{EFcoords}\ee
where
\be
 A(\bar{r}) = \frac{A_2}{\bar{r}^2} + \frac{A_1}{\bar{r}} + A_0, \qquad B(\bar{r}) =
\frac{B_1}{\bar{r}} + B_0,
\ee
and $A_i$, $B_i$ are determined by requiring that the solution in the
coordinates $(v,\bar{r},x,\phi',\psi')$ is analytic at
$\bar{r}=0$. First note that the scalars $X^I$ are already analytic at
$\bar{r}=0$. The electromagnetic potentials $A^i$ are given by
\ba
\label{eqn:gaugehorizon}
A^i &=& \frac{4 q_i \bar{r}^2}{q^3} \left( 1 + {\cal O}(\bar{r}) \right) dv
- \frac{q_i}{2} \left( 1 + x + {\cal O}(\bar{r}) \right) d\phi' -
\frac{q_i}{2} \left( 1 - x - \frac{2 \mathcal{Q}_i q_i - \sum_j
\mathcal{Q}_j q_j}{q^3} +
{\cal O}(\bar{r}) \right) d\psi' \nn &-& \left[\frac{b^i_0 B_1}{\bar{r}} +
\frac{4 q_i A_2}{q^3} + b^i_0 B_0 + b^i_{-1} B_1 + {\cal O}(\bar{r})
\right] d\bar{r},
\ea
where $b^i_0$ and $b^i_{-1}$ are certain constants and ${\cal O}(\bar{r})$
denotes terms that can be expanded as a series in positive powers of
$\bar{r}$.
This is analytic up to a term that can be removed by a gauge transformation.
For the metric, we find that
that $g_{\bar{r}\psi'}$ diverges as $1/\bar{r}$ unless we choose $A_2 = -L^2(q_1
q_2 q_3)^{1/3} B_1/(2R)$, where $L$ was introduced in \reef{bigL}. We then
find that $g_{\bar{r}\bar{r}}$ has a $1/\bar{r}^2$
divergence unless we choose\footnote{
The overall sign here is arbitrary; making the opposite choice would
lead to an extension of the metric through the past horizon rather than
the future horizon.}
\be
 B_1 = -\frac{q}{2 L}.
\ee
This implies
\be
 A_2 = \frac{L q^2}{4R}.
\ee
Now $g_{\bar{r}\bar{r}}$ diverges as $1/\bar{r}$ unless we choose
\be
 A_1 = \frac{1}{4R^2 L q^2} \left[\mathcal{Q}_1 \mathcal{Q}_2 \mathcal{Q}_3 - R^2
\sum_i \mathcal{Q}_i q_i^2 + q^3 (q_1 + q_2 + q_3) R^2 \right].
\ee
The metric is now analytic at $\bar{r}=0$. However we still have the freedom
to choose the finite part of the coordinate transformation. After the
above transformation, $g_{\bar{r}\bar{r}}$ is a linear function of $x$ at $\bar{r}=0$
and we can choose $A_0$, $B_0$ to cancel this function so that
$g_{\bar{r}\bar{r}}=0$ at $\bar{r}=0$. The expressions for $A_0$ and $B_0$ are lengthy
and unilluminating so we shall not present them here. The metric
finally takes the form
\ba
 ds_5^2 &=& -\frac{16 \bar{r}^4}{q^4} dv^2 + 2 \frac{R}{L} dv
d\bar{r} + \frac{4 \bar{r}^3 \sin^2 \bar{\theta}}{R q} dv d\phi' +
\frac{4 \bar{r} R}{q} dv d\psi' \nn
&+& \frac{1}{L} (q_1 + q_2
+ q_3) \bar{r} \sin^2 \bar{\theta} d\bar{r} d\phi' + 2 \left[\frac{qL}{2R}
\cos \bar{\theta} - c \right] d\bar{r} d\psi' \\
&+& L^2 d{\psi'}^2 + \frac{q^2}{4} \left[ d\bar{\theta}^2 +
\sin^2 \bar{\theta}
\left( d\phi' - d\psi' \right)^2 \right] + \ldots \nonumber
\ea
We have set $x = \cos \bar{\theta}$. The ellipsis denotes terms in
$g_{\bar{r}\bar{r}}$ starting at ${\cal O}(\bar{r})$, as well as
subleading (integer) powers of $\bar{r}$ in all of the metric components
explicitly written above. The constant $c$ is given by
\be
 c = \frac{1}{2LR q_1 q_2 q_3} \left[ \mathcal{Q}_1 \mathcal{Q}_2 \mathcal{Q}_3 - R^2 \sum_{i < j}
(\mathcal{Q}_i + \mathcal{Q}_j) q_i q_j - q^3 (q_1 + q_2 + q_3) R^2 \right].
\ee
The above metric is analytic in $\bar{r}$ hence so is its determinant.
At $\bar{r}=0$, the determinant vanishes
if, and only if, $\sin^2 \bar{\theta} = 0$, which is just a coordinate
singularity. Hence the inverse metric is also analytic in $\bar{r}$ so the
above coordinates define an analytic extension of our solution through
the surface $\bar{r}=0$.

The supersymmetric Killing vector field $V = \partial/\partial v$ is
null at $\bar{r}=0$. Furthermore, $V_\mu dx^\mu = (R/L) d\bar{r}$ at $\bar{r}=0$ so $V$
is normal to the surface $\bar{r}=0$. Hence $\bar{r}=0$ is a null hypersurface and
a Killing horizon of $V$, i.e., our solution has an event horizon at
$\bar{r}=0$.

The metric of a spatial cross-section of the horizon can be written
\beq
  ds_\rom{horizon}^2 = L^2 d{\psi'}^2
  +\frac{q^2}{4}
   \Big(d\bar{\theta}^2 + \sin^2{\bar{\theta}} d\chi^2\Big)
\eeq
with $\chi= \phi' - \psi'$.
The near-horizon geometry is locally $AdS_3 \times S^2$ where the $AdS_3$ has
radius $(q_1 q_2 q_3)^{1/3}$ and the $S^2$ has radius $(q_1 q_2
q_3)^{1/3}/2$.

\subsection{The IIB solution}

Consider now the IIB solution \reef{eqn:2bmetric}. As $y\to-\infty$, the
conformal factors multiplying the three terms in \reef{eqn:2bmetric}
remain finite and non-zero.
Hence, after transforming to the coordinates
$(v,\bar{r},\bar{\theta},\phi',\psi')$, the only part of the metric
that is not manifestly regular at $\bar{r}=0$ is the part involving
$A^3$. To make this regular we need a gauge transformation, \ie a
shift in $z$. Using equation \reef{eqn:gaugehorizon}, the required
shift is
\be
dz = dz' + \left[\frac{b^3_0 B_1}{\bar{r}} + \frac{4 A_2}{q_1 q_2} + b^3_0 B_0
+ b^3_{-1} B_1 \right] d\bar{r},
\ee
which gives
\ba
dz+A^3 &=& dz' + \frac{4 \bar{r}^2}{q_1 q_2} \left( 1 + {\cal
O}(\bar{r}) \right) dv - \frac{q_3}{2} \left( 1 + \cos \bar{\theta} +
{\cal O}(\bar{r}) \right) d\phi' \nn &-& \frac{q_3}{2} \left( 1 - \cos
\bar{\theta} - \frac{2 \mathcal{Q}_3 q_3 - \sum_j \mathcal{Q}_j
q_j}{q^3} + {\cal O}(\bar{r}) \right) d\psi' + {\cal O}(\bar{r})
d\bar{r}.
\ea
This is manifestly analytic at $\bar{r}=0$. We still have $V_\mu dx^\mu
\propto d\bar{r}$ at $\bar{r}=0$ where $V = \partial/\partial v$ so $\bar{r}=0$ is a
Killing horizon of $V$.

\subsection{Null orbifold singularity of rings with $L=0$ and $q_i\neq
0$}
\label{app:nullorb}

Consider now solutions where all the dipoles $q_i$ are non-zero but
$L=0$. The analysis is similar to the case $L>0$ so we shall simply
sketch it. Let $y=-R^2/\bar{r}^2$ and change as in \reef{EFcoords}. Take
$A=A_2/\bar{r}^2$ and $B=B_2/\bar{r}^2$. $A_2$ and $B_2$ can be chosen
to cancel the divergent term in $g_{\bar{r}\bar{r}}$. There are no other
divergences in $g_{\bar{r}\bar{r}}$ nor $g_{\bar{r}\psi'}$. The
determinant of the metric vanishes at $\bar{r}=0$, but this comes from
the coefficient $\bar{r}^2$ in $g_{\psi'\psi'}$ and is a coordinate
singularity analogous to the one in the Poincare patch of AdS$_3$.
There remains a $\bar{r}^0$ piece in $g_{\bar{r}\bar{r}}$ which is $x$
dependent (but does not vanish at any $x$), and could be eliminated
by adding an $\bar{r}^0$ term to $B$ which is also a linear function of
$x$, but this is not actually necessary. So there is no curvature
singularity at $y=-\infty$. The near-horizon (or, more
appropriately, ``near-core") geometry is more simply expressed in
coordinates which only cover the outer region. Changing
$y=-R^2/(\epsilon\tilde r^2)$ and $t=\tilde t/\epsilon$, and sending
$\epsilon\to 0$ we find
\be
ds^2_\mathit{5}=\frac{4\tilde r^2}{q}\;d\tilde td\psi+q^2\frac{d\tilde
r^2}{\tilde r^2} +\frac{q^2}{4}(d\tilde\theta^2+\sin^2\tilde\theta
d\chi^2)\,.
\ee
This is, like in the cases with $L>0$, locally AdS$_3\times S^2$. The AdS$_3$
part is written in double-null form, and since the orbits of
$\partial_\psi$ are closed we find a null orbifold singularity at
$\tilde r=0$ instead of a regular horizon. The near-core limit for the
corresponding IIB solutions is easily obtained by comparing to
\reef{nearh}.

\medskip

The solutions where some of the $q_i$ vanish, and possibly also some
$Q_i$, are studied in Section~\ref{sec:cases}.

\end{document}